\documentclass[iop]{emulateapj}
\usepackage{amssymb}
\usepackage{amsmath}
\usepackage{rotating}
\usepackage{longtable}

\newcommand{\msun}{M_\odot}
\newcommand{\lsun}{L_\odot}

\begin{document}
\title{Variable Stars in Large Magellanic Cloud Globular Clusters I: NGC 1466}
\author{Charles A. Kuehn\altaffilmark{1}, Horace A. Smith\altaffilmark{1}, M\'arcio Catelan\altaffilmark{3}, Barton J. Pritzl\altaffilmark{4}, Nathan De Lee\altaffilmark{1,5}, Jura Borissova\altaffilmark{6}}
\altaffiltext{1}{Department of Physics and Astronomy, Michigan State University, East Lansing, MI 48824, USA; kuehncha@msu.edu, smith@pa.msu.edu}
\altaffiltext{3}{Pontificia Universidad Cat$\rm{\acute{o}}$lica de Chile, Departamento de Astronomi\'ia y Astrof\'isica, Santiago, Chile; mcatelan@astro.puc.cl}
\altaffiltext{4}{Department of Physics and Astronomy, University of Wisconsin Oskosh, Oshkosh, WI 54901, USA; pritzlb@uwosh.edu}
\altaffiltext{5}{Current address: Department of Astronomy, University of Florida, Gainesville, FL 32611, USA; nedelee@astro.ufl.edu}
\altaffiltext{6}{Departamento de F\'isica y Astronom\'ia, Falcultad de Ciencias, Universidad de Valpara\'iso, Valpara\'iso, Chile; jura.borissova@uv.cl}

\begin{abstract}
This is the first in a series of papers studying the variable stars in Large Magellanic Cloud globular clusters.  The primary goal of this series is to better understand how the RR Lyrae stars in Oosterhoff-intermediate systems compare to those in Oosterhoff I/II systems.  In this paper we present the results of our new time-series $BV$ photometric study of NGC 1466.  A total of $62$ variables were identified in the cluster, of which $16$ are new discoveries.  The variables include $30$ RRab stars, $11$ RRc's, $8$ RRd's, $1$ candidate RR Lyrae, $2$ long-period variables, $1$ potential anomalous Cepheid, and $9$ variables of undetermined classification.  We present photometric parameters for these variables.

For the RR Lyrae stars physical properties derived from Fourier analysis of their light curves are presented.  The RR Lyrae stars were used to determine a reddening-corrected distance modulus of $(m-M)_{0} = 18.43\pm 0.15$.  We discuss several different indicators of Oosterhoff type and find NGC 1466 to be an Oosterhoff-intermediate object.

\end{abstract}


\section{Introduction}
Studies of RR Lyrae stars in the Milky Way globular clusters reveal what is traditionally known as the Oosterhoff dichotomy; the tendency for these clusters to be either Oosterhoff I (Oo-I) or Oosterhoff II (Oo-II) objects with a relatively clear zone of avoidance between these two groups.  The left panel of Figure 5 in \citet{mc09} strikingly illustrates the Oosterhoff dichotomy in a plot of the average period of the RR Lyrae stars of Bailey type ab (RRab) versus cluster metallicity for Milky Way globular clusters.  When one looks at the nearby dwarf galaxies and their globular clusters, one does not see an Oosterhoff dichotomy as these objects fall not only into the Oo-I/II groups but also into the gap between those two groups.  In fact the distribution of these extragalactic objects seems to peak in the zone of avoidance \citep{mc09}.

The existence of Oosterhoff-intermediate objects (Oo-int), objects that fall into the zone of avoidance, in the nearby dwarf galaxies poses a significant challenge to the hierarchical merger model for the formation of the Milky Way halo.  The lack of Oo-int clusters in the halo means that the objects that were accreted to form the halo could not have had properties entirely like the present-day nearby dwarf galaxies.

This is the first of a series of papers focusing on variable stars in globular clusters in the Large Magellanic Cloud (LMC).  The goal of this series is to better understand how the RR Lyrae stars in Oo-int systems compare to those in Oo-I/II systems.  The LMC is an ideal place for this study because of the large number of Oo-int clusters that it contains \citep{bo94,mc09}; of the twelve LMC globular clusters that contain at least $5$ RRab variables, five of these clusters are Oo-int (NGC 1466, NGC 1853, NGC 2019, NGC 2210, and NGC 2257).  In this paper we identify, classify, and discuss the variable stars found in the LMC globular cluster NGC 1466.  Subsequent papers will focus on the variable stars in other LMC globular clusters.

NGC 1466 is an old globular cluster that is located relatively far from the center of the LMC.  It has a metal abundance of [Fe/H] $\approx -1.60$, see discussion in section \ref{sec:abprop}, and is not very reddened, E(B-V) = $0.09\pm0.02$ \citep{wa92b}.  \citet{jj99} used color-magnitude diagrams obtained with the Hubble Space Telescope to determine that the age of NGC 1466 is within $1$ Gyr of the ages of the Milky Way globular clusters M3 (NGC 5272) and M92 (NGC 6341).

NGC 1466 features a well-populated horizontal branch that extends through the instability strip (see Figure 5 in Walker 1992); thus it is expected to contain a significant number of RR Lyrae stars. RR Lyrae stars were first found in NGC 1466 by \citet{tw53}, and additional RR Lyrae were discovered by \citet{we71}.  The most recent study of variables in NGC 1466 was conducted by \citet{wa92b}, who found $42$ RR Lyrae stars; $25$ RRab and $17$ RRc stars.  Due to the density of unresolved stars in the cluster core, Walker did not perform photometry of stars within a radius of $13$ arcsec from the cluster center, suggesting that there are probably additional RR Lyrae stars that could not be detected in his study.  The core radius of NGC 1466 is $10.7\pm0.4$ arcsec, as measured by \citet{mg03} using images from the Hubble Space Telescope, which is entirely within the region of the cluster where Walker was unable to perform photometry.  Advances in instrument resolution and image-subtraction techniques now make it easier for us to search for variable stars in more crowded regions.

In this paper we present the results of a new search for variable stars in NGC 1466 and use the RR Lyrae in NGC 1466 to confirm the distance modulus and Oosterhoff classification of the cluster.  We also present Fourier fits to the light curves and derive physical properties for the stars from these fits.

\section{Observations and Data Reduction}

A total of $40$ $V$ and $37$ $B$ images of NGC 1466 were obtained using the SOI imager on the SOAR $4$-m telescope in February and December of 2008. An additional $45$ $V$ and $43$ $B$ images were obtained using ANDICAM on the SMARTS $1.3$-m telescope operated by the SMARTS consortium\footnote[7]{www.astro.yale.edu/smarts} from September 2006 to January 2007.   This represents a larger data set than the 35 BV pairs obtained by \citet{wa92b}.  Exposure times for the SOAR observations were between $30$s and $300$s  for $V$ and between $60$s and $300$s for $B$, but were usually $120$s and $180$s for the $V$ and $B$ observations, respectively.  SMARTS exposures were $450$s for each filter.

The images from both telescopes were bias subtracted and flat-field corrected using IRAF\footnote[8]{IRAF is distributed by the National Optical Astronomical Observatory, which is operated by the Association of Universities for Research in Astronomy, Inc., under cooperative agreement with the National Science Foundation.}.  As noted in \citet{wa92b}, NGC 1466 is located near two very bright stars which create problems with scattered light across some of the images, especially in many of the images obtained with the SMARTS telescope.  We used the procedure described in detail by \citet{sh88} to remove this scattered light; this was the same method employed by \citet{wa92b}.  Peter Stetson's Daophot II/Allstar packages \citep{st87,st92,st94} were used to locate, measure, and subtract stars from all images; this produced images that only contained the sky background and scattered light.  The subtracted images were then smoothed using a $40$x$40$ pixel median filter.  The average sky level in the smoothed images was determined and then the smoothed images were subtracted from the original images, removing the background light.  The average sky values were then added back into the new images, creating a final set of images with a constant sky background.

Stetson's Daophot II/Allstar packages were run on the new set of images in to order obtain instrumental magnitudes for each star.  Observations of the Landolt standard fields PG0231, RU149, SA95, and SA98 \citep{la92} were used to determine an initial transformation from instrumental magnitudes to the standard system. \citet{jj99} showed Walker's photometry to be in good agreement with HST/WFPC2 observations of NGC 1466, after these observations were transformed from the on-board HST system to the standard Johnson-Cousins system.  We compared the results of our initial transformation to the standard system using Walker's photometry of isolated non-variable stars across a range of brightnesses within the cluster.  We made small corrections to the zero points and color terms from our initial transformation equations in order to achieve better agreement with Walker's photometry.  This produced an average difference between our data and Walker's of $ \Delta V = 0.012\pm0.01$ and $\Delta B =0.026\pm0.02$.  This is a similar level of scatter to that obtained by \citet{jj99} in their comparison with Walker.

The profile fitting photometry used in Daophot II/Allstar does not work well in very crowded regions, such as the center of NGC 1466.  In order to locate variable stars in this region, the SOAR data on the cluster center was also searched using the ISISv2.2 image subtraction program \citep{ca00}.  The original images were used for the ISIS processing as the scattered light is less of a problem when using image subtraction techniques.  The light curves produced by ISIS are in relative fluxes.  There is a method to transform the relative fluxes obtained by ISIS into magnitudes in the standard system by using Daophot to obtain magnitudes of the stars in the reference frame created by ISIS \citep{mo01}.  Four of our variable stars were found only by ISIS (V$31$, V$51$, V$54$, V$60$).  These stars  are located deep in the cluster center and are too crowded to obtain accurate profile fitting photometry, even in the reference image.  These four light curves are shown in relative fluxes; for the other variables we present their light curves obtained from Daophot II/Allstar.

\begin{figure*}[t]
\includegraphics[width=0.49\textwidth]{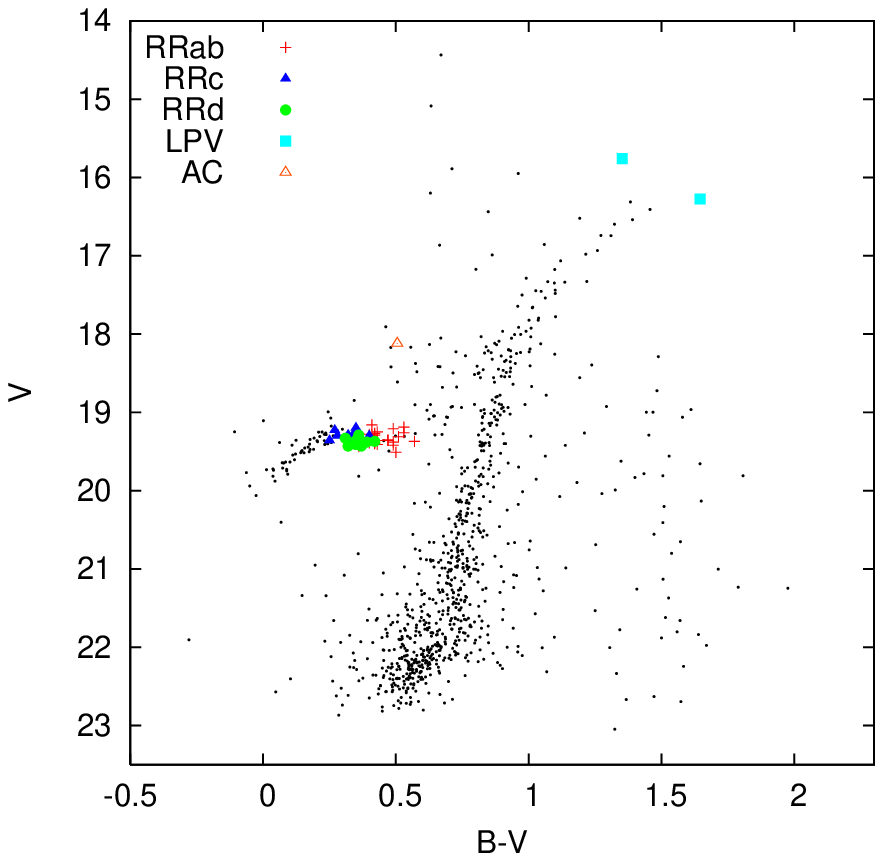}
\includegraphics[width=0.49\textwidth]{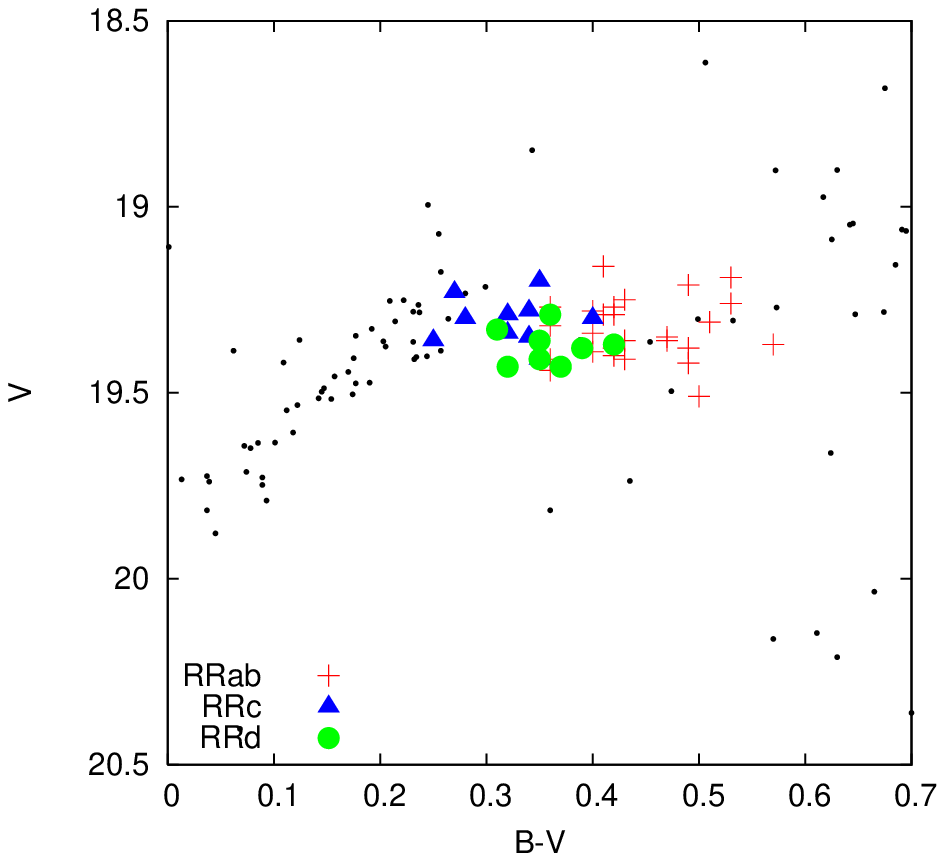}
\caption{$V$,$B-V$ CMD for NGC 1466 (full CMD on left) with the position of the RR Lyrae variables also indicated.  Red plus symbols indicate RRab stars, dark blue filled triangles indicate RRc's, green circles indicate RRd's, and light blue squares indicate long-period variables.  The potential Anomalous Cepheid is indicated by an open red triangle.  A zoomed-in view of the CMD showing the HB is displayed on the right.  Stars within the central $5$ arcseconds of the cluster are not plotted due to photometric error from blending.}
\label{cmd}
\end{figure*}

Figure \ref{cmd} shows our CMD for NGC 1466.  The cluster features a well-defined horizontal branch (HB) that extends across the instability strip.  The majority of the HB stars are blue, with few HB stars located redward of the instability strip.  As was noted by \citet{wa92b}, the RR Lyrae stars appear to be less luminous than the blue HB stars near the blue edge of the instability strip.  As can be seen in the figure, there are some non-variable stars in the instability strip region.  These are due to either poor photometry due to blending in the cluster center or are line-of-sight field stars.

\section{Variable Stars}
 We used the $V$-band time series, which had more phase points than the $B$-band one, to identify candidate variable stars using Peter Stetson's Allframe/Trial package \citep{st94}.  Period searches on the candidate variables were carried out using Supersmoother \citep{re94}, {\tt period04} \citep{le04}, and a discrete Fourier transform \citep{ferraz81} as implemented in the Peranso software suite\footnote[9]{http://www.peranso.com/}.  This produced light curves which were then examined by eye to confirm the reality of each variable star candidate.  Light curves of potential RR Lyrae stars were then fit to template light curves \citep{ly98} and checked by eye in order to confirm variable classification and period.  Resulting primary periods are typically good to $\pm 0.00001$ or $0.00002$ days while errors for the secondary periods of RRd stars are slightly worse but on the same order of magnitude.  A similar search of the light curves produced with ISIS was done to find additional variables.

\begin{figure*}
\begin{center}
\includegraphics[width=0.45\textwidth]{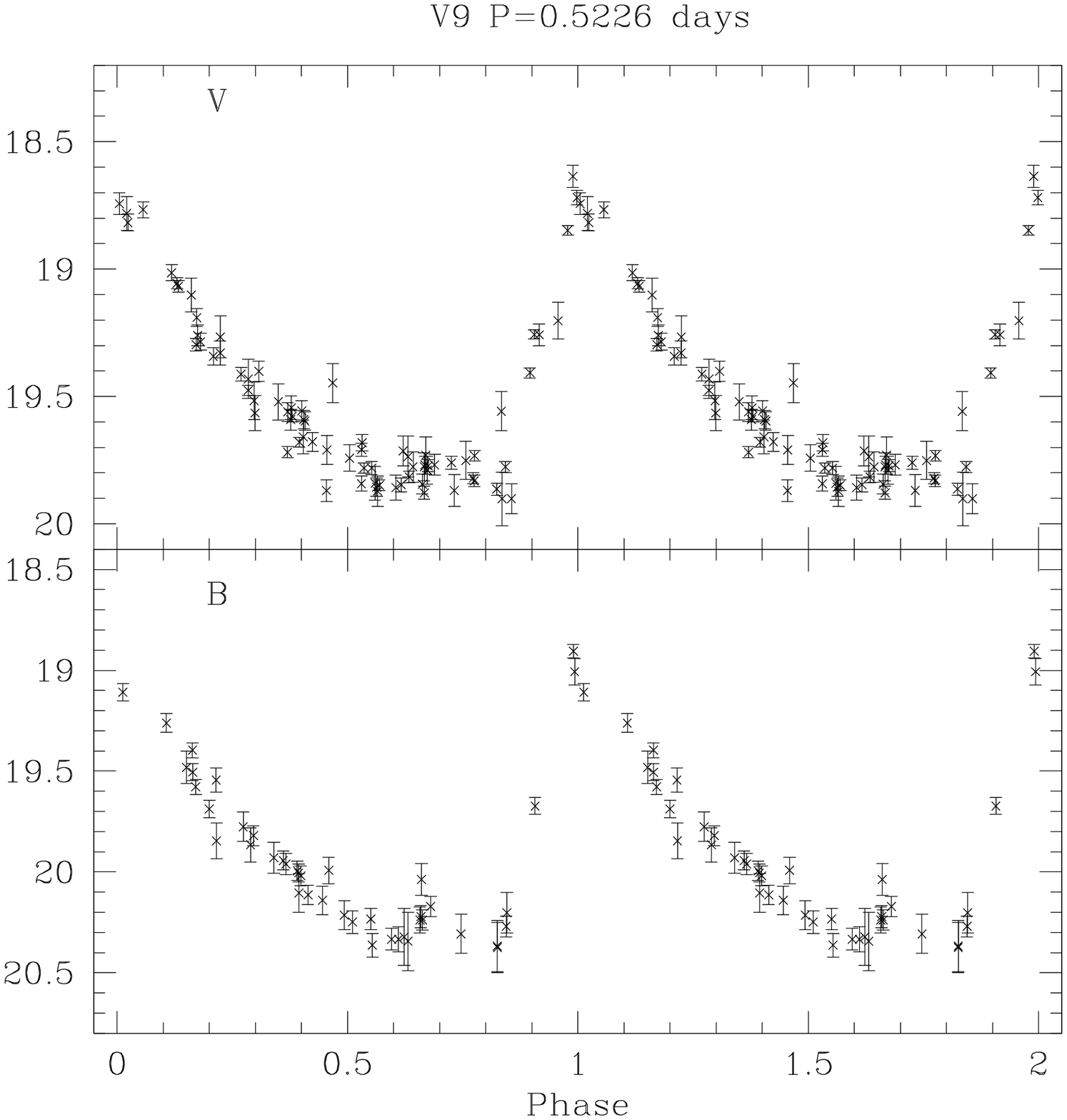}
\includegraphics[width=0.45\textwidth]{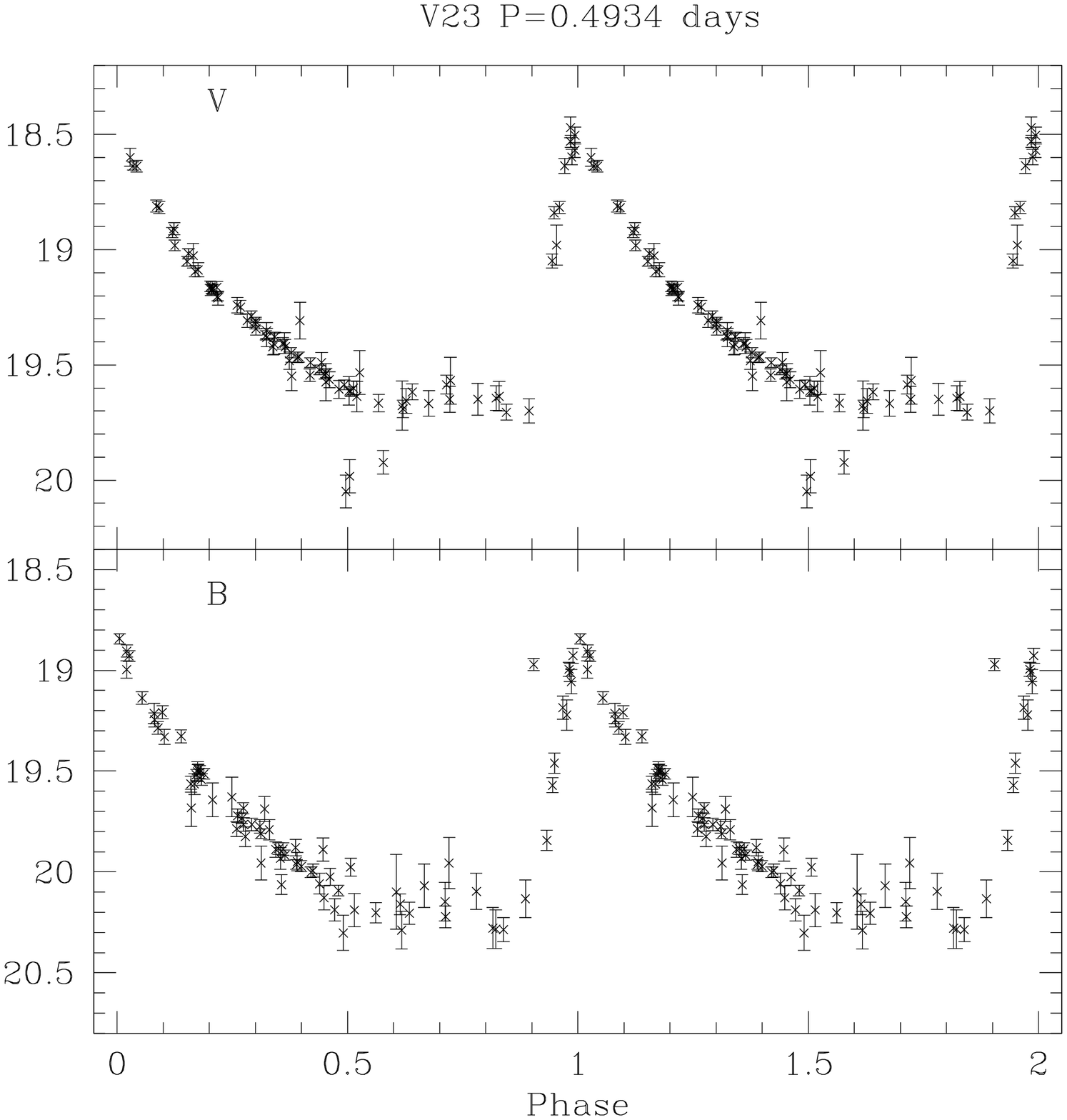}
\caption{Sample light curves for the RRab stars in NGC 1466.  (The full set of light curves can be found in the electronic version of this paper.)}
\label{abcurves}
\end{center}
\end{figure*}

\begin{figure*}
\begin{center}
\includegraphics[width=0.45\textwidth]{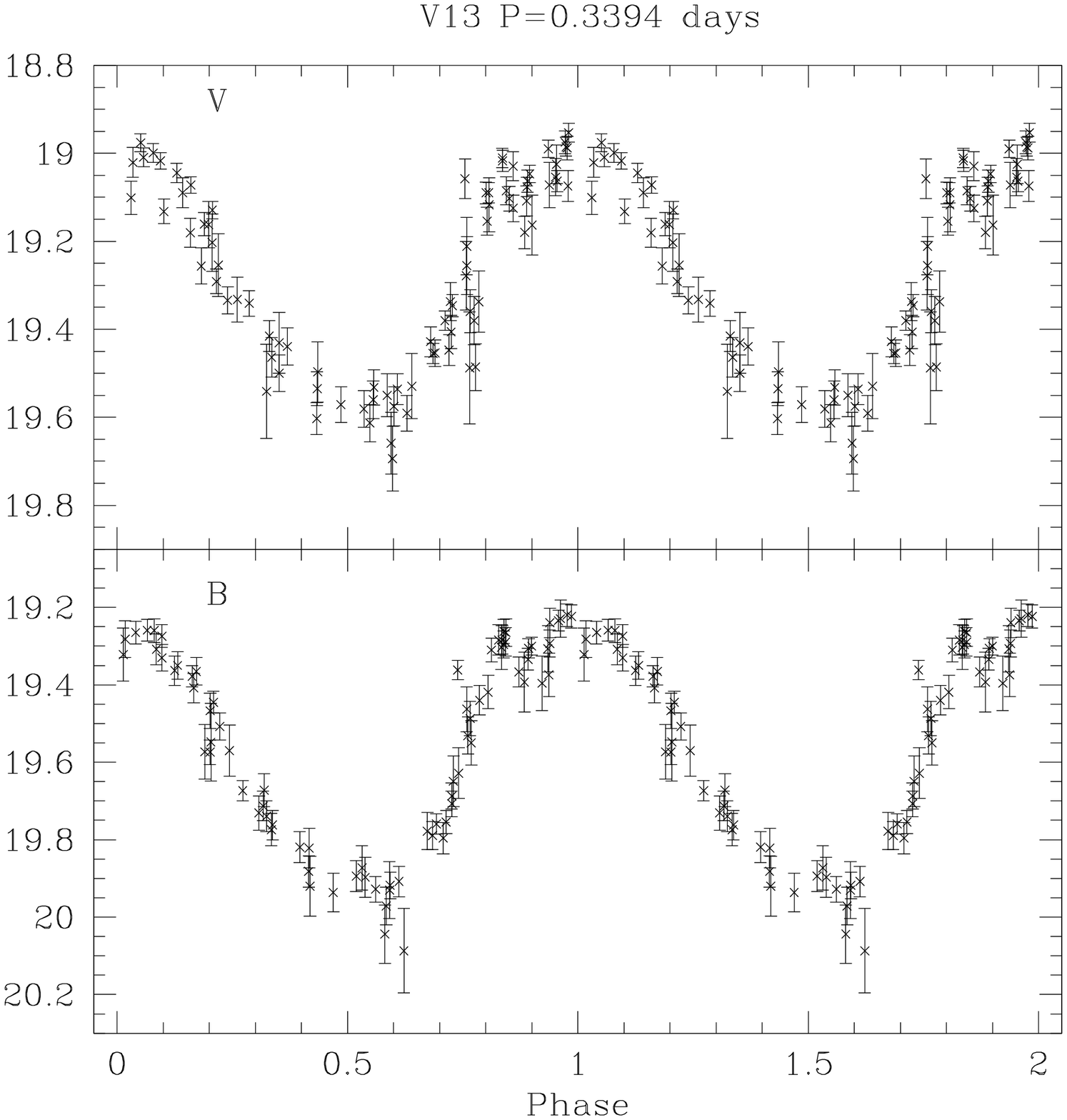}
\includegraphics[width=0.45\textwidth]{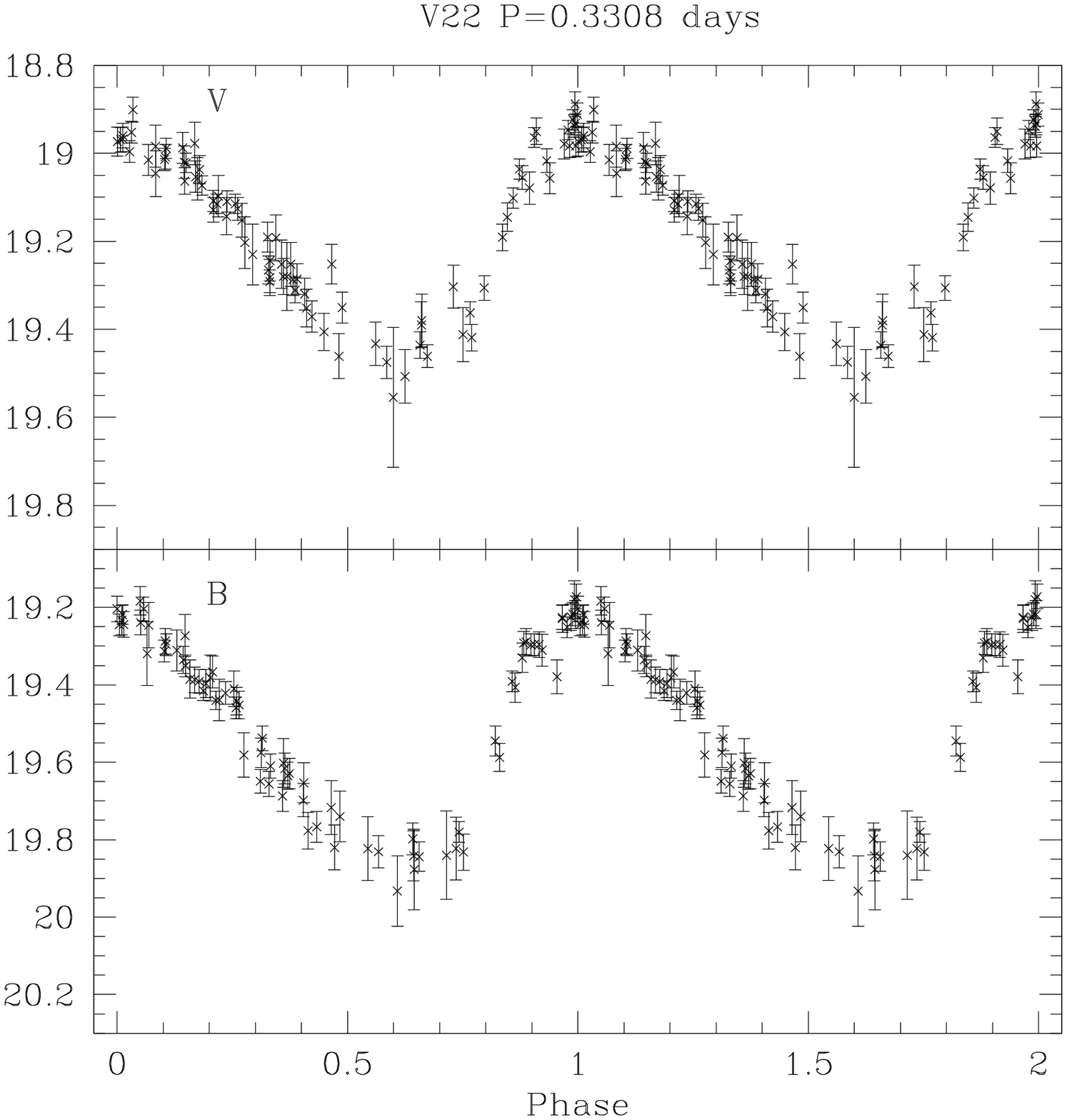}
\caption{Sample light curves for the RRc stars in NGC 1466.  (The full set of light curves can be found in the electronic version of this paper.)}
\label{ccurves}
\end{center}
\end{figure*}

\begin{deluxetable*}{lccccccccccc}
\tablewidth{0pc}
\tabletypesize{\scriptsize}
\tablecaption{Photometric Parameters for Variables in NGC 1466, Excluding RRd Stars}
\tablehead{\colhead{ID}& \colhead{RA (J2000)} & \colhead{DEC (J2000)} & \colhead{Type}& \colhead{$P$ (days)}& \colhead{$A_{V}$}& \colhead{$A_{B}$}& \colhead{$\langle V\rangle$}& \colhead{$\langle B\rangle$}& \colhead{$(B-V)_{mag}$} & \colhead{Other IDs}&\colhead{Max. Light (HJD)}}
\startdata
V1& 03:44:44.7 & $-$71:40:44.9 & RRab-BL& 0.55035& 0.64& 0.77& 19.380& 19.793& 0.421& Wa415& 2454094.6963\\
V4& 03:44:20.3 & $-$71:39:57.1 & RRab-BL& 0.57239& 0.87& 1.16& 19.406& 19.804& 0.425& Wa67& 2454094.7100\\
V5& 03:44:26.6 & $-$71:39:48.9 & RRab& 0.57863& 0.67& 0.82& 19.391& 19.781& 0.403& Wa108& 2454510.5801\\
V6& 03:44:30.8 & $-$71:39:54.8 & RRab-BL& 0.52805& 1.19& 1.49& 19.290& 19.672& 0.410& Wa182& 2454502.2325\\
V8& 03:44:26.3 & $-$71:40:06.5 & RRab-BL& 0.51970& 0.89& 1.2& 19.273& 19.599& 0.358& Wa106& 2454094.7502\\
V9& 03:44:26.7 & $-$71:40:17.2 & RRab& 0.52260& 1.24& 1.52& 19.414& 19.743& 0.362& Wa109& 2454086.9323\\
V12& 03:44:28.9 & $-$71:40:15.0 & RRab& & & & & & & Wa146& \\
V13& 03:44:33.8 & $-$71:40:48.1 & RRc& 0.33936& 0.55& 0.69& 19.298& 19.564& 0.276& Wa262& 2454094.5656\\
V14& 03:44:36.2 & $-$71:40:48.7 & RRab-BL& 0.56414& 0.93& 1.29& 19.402& 19.803& 0.421& Wa320& 2454501.3551\\
V15& 03:44:35.4 & $-$71:40:44.2 & RRab& 0.56708& 0.81& 1.05& 19.209& 19.675& 0.487& Wa303& 2454094.9557\\
V16& 03:44:34.9 & $-$71:40:45.1 & RRab& 0.69427& 0.53& 0.69& 19.360& 19.817& 0.466& Wa290& 2454092.6208\\
V18& 03:44:26.7 & $-$71:40:14.5 & ? & 0.48837 & 0.15& 0.29& 18.903& 20.168& 1.268& Wa110& 2454510.7987 \\
V19& 03:44:47.1 & $-$71:40:39.8 & RRab& 0.61620& 0.61& 0.79& 19.360& 19.785& 0.435& Wa426& 2454097.6522\\
V20& 03:44:30.7 & $-$71:41:22.5 & RRab& 0.59084& 1.05& 1.33& 19.280& 19.648& 0.396& Wa183& 2454094.6908\\
V21& 03:44:51.9 & $-$71:39:53.4 & RRab& 0.52150& 1.22& 1.54& 19.315& 19.632& 0.359& Wa448& 2454048.0181\\
V22& 03:44:35.5 & $-$71:41:05.7 & RRc& 0.33081& 0.51& 0.65& 19.200& 19.543& 0.353& Wa301& 2454094.7036\\
V23& 03:44:38.6 & $-$71:41:24.7 & RRab& 0.49342& 1.25& 1.51& 19.291& 19.681& 0.421& Wa361& 2454094.4714\\
V26& 03:44:34.2 & $-$71:40:51.4 & RRab& 0.51364& 1.03& 1.38& 19.339& 19.702& 0.400& Wa274& 2454094.7460\\
V28& 03:44:21.3 & $-$71:40:38.1 & RRc& 0.32039& 0.57& 0.7& 19.338& 19.652& 0.324& Wa71& 2454094.7532\\
V30& 03:44:31.3 & $-$71:39:52.0 & RRab& 0.69445& 0.66& 0.91& 19.273& 19.677& 0.421& Wa194& 2454095.2544\\
V31& 03:44:32.1 & $-$71:39:50.5 & ? & 0.34272& & & & & & & 2454510.6763\\
V32& 03:44:29.4 & $-$71:40:13.7 & AC?& 0.49661& 0.42& 0.70& 18.118& 18.614& 0.505& & 2454094.7437\\
V33& 03:44:29.0 & $-$71:40:43.9 & RRc& 0.27841& 0.51& 0.57& 19.355& 19.600& 0.247& Wa148& 2454094.8536\\
V34& 03:44:28.0 & $-$71:39:60.0 & ? & 0.61594& 0.31& 0.45& 18.194& 18.896& 0.709& Wa129& 2454040.9638\\
V35& 03:44:29.7 & $-$71:40:42.5 & RRab-BL& 0.58172& 0.6& 0.78& 19.248& 19.668& 0.432& Wa164& 2454094.6090\\
V38& 03:44:49.0 & $-$71:40:48.7 & RRc& 0.34858& 0.44& 0.54& 19.354& 19.682& 0.335& Wa437& 2454092.5444\\
V39& 03:44:30.6 & $-$71:40:07.2 & RRab& 0.56248& 1.12& 1.37& 19.156& 19.538& 0.407& Wa179& 2454510.8795\\
V40& 03:44:31.0 & $-$71:40:01.1 & RRab& 0.53606& & & & & & Wa185& 2454092.8046\\
V41& 03:44:39.4 & $-$71:40:21.3 & RR?& & & 0.59& & 19.598& & & \\
V42& 03:44:30.4 & $-$71:41:00.6 & RRab& 0.63482& 0.51& 0.66& 19.384& 19.865& 0.489& Wa177& 2454094.6637\\
V44& 03:44:22.0 & $-$71:38:13.1 & RRab& & & & & & & Wa75,S27& \\
V45& 03:44:23.1 & $-$71:41:16.0 & RRc& 0.31043& 0.38& 0.49& 19.233& 19.503& 0.275& Wa82,S27& 2454094.4913\\
V46& 03:44:27.4 & $-$71:39:14.0 & RRab& 0.68909& 0.38& 0.49& 19.352& 19.813& 0.466& Wa121& 2454094.3662\\
V48& 03:44:29.4 & $-$71:40:00.0 & RRc& 0.37697& 0.44& 0.54& 19.288& 19.580& 0.316& Wa153,S3& 2454092.6531\\
V49& 03:44:30.2 & $-$71:40:09.5 & RRab& 0.68635& 0.53& 0.7& 19.255& 19.773& 0.528& & 2454510.2159\\
V50& 03:44:31.1 & $-$71:40:28.7 & RRab-BL& 0.58194& 1.05& 1.24& 19.444& 19.778& 0.356& & 2454092.5885\\
V51& 03:44:31.6 & $-$71:40:12.7 & ? & 0.35892& & & & & & & 2454410.5428\\
V52& 03:44:31.6 & $-$71:40:34.6 & RRc& & & & & & & & \\
V53& 03:44:31.9 & $-$71:40:15.8 & LP & & & & & & & & \\
V54& 03:44:32.1 & $-$71:40:12.9 & ? & 0.55608 & & & & & & & 2454510.3605\\
V55& 03:44:32.5 & $-$71:40:13.3 & RRab-BL& 0.60951& & & & & & & 2454510.6458\\
V56& 03:44:33.4 & $-$71:40:01.8 &RRab& 0.62238& 0.73& 0.96& 19.189& 19.701& 0.527& & 2454510.5354\\
V57& 03:44:33.5 & $-$71:39:51.7 & RRc& 0.35692& 0.42& 0.5& 19.295& 19.685& 0.395& Wa249,S29& 2454094.6595\\
V58& 03:44:33.5 & $-$71:39:59.0 & RRab& 0.56752& 1.05& 1.28& 19.368& 19.905& 0.570& & 2454510.4638\\
V59& 03:44:33.7 & $-$71:40:20.2 & ? & &0.46& 0.54& 18.194& 18.651& 0.460&  & 2454511.0736\\
V60& 03:44:33.9 & $-$71:40:11.4 & AC?,RRab? & 0.52210 & & & & & & & 2454510.1325\\
V61& 03:44:33.9 & $-$71:40:01.7 & RRab-BL& 0.66927& 0.86& 1.13& 19.510& 19.985& 0.497& & 2454510.6914\\
V62& 03:44:35.5 & $-$71:40:35.8 & ? & 0.50414& 0.71& 1.08& 18.758& 19.266& 0.531& & 2454094.2817\\
V63& 03:44:35.6 & $-$71:39:54.0 & RRab& 0.58760& 0.82& 0.99& 19.415& 19.890& 0.490& Wa306,S31& 2454510.6730\\
V64& 03:44:38.4 & $-$71:40:39.5 & RRab& 0.68767& 0.49& 0.59& 19.307& 19.809& 0.508& Wa356,S30& 2454092.8083\\
V65& 03:44:39.0 & $-$71:40:03.7 & LP & & & & & & & & \\
V66& 03:44:40.0 & $-$71:41:10.3 & RRc& 0.33633& 0.52& 0.67& 19.277& 19.608& 0.342& Wa382,S17& 2454094.7809\\
V67& 03:44:45.1 & $-$71:42:13.1 & ? & & & & & & & & \\
V68& 03:44:56.5 & $-$71:38:54.3 & RRc& 0.34914& 0.47& 0.65& 19.410& 19.749& 0.348& Wa458& 2454040.8262\\
\enddata
\tablecomments{The RRab stars that potentially exhibit the Blazhko effect are indicated as `RRab-BL', see section 3.1 for further discussion on the detection of the Blazhko effect in the RRab stars in NGC 1466.}
\label{vartable}
\end{deluxetable*}

\begin{figure*}
\begin{center}
\includegraphics[width=0.45\textwidth]{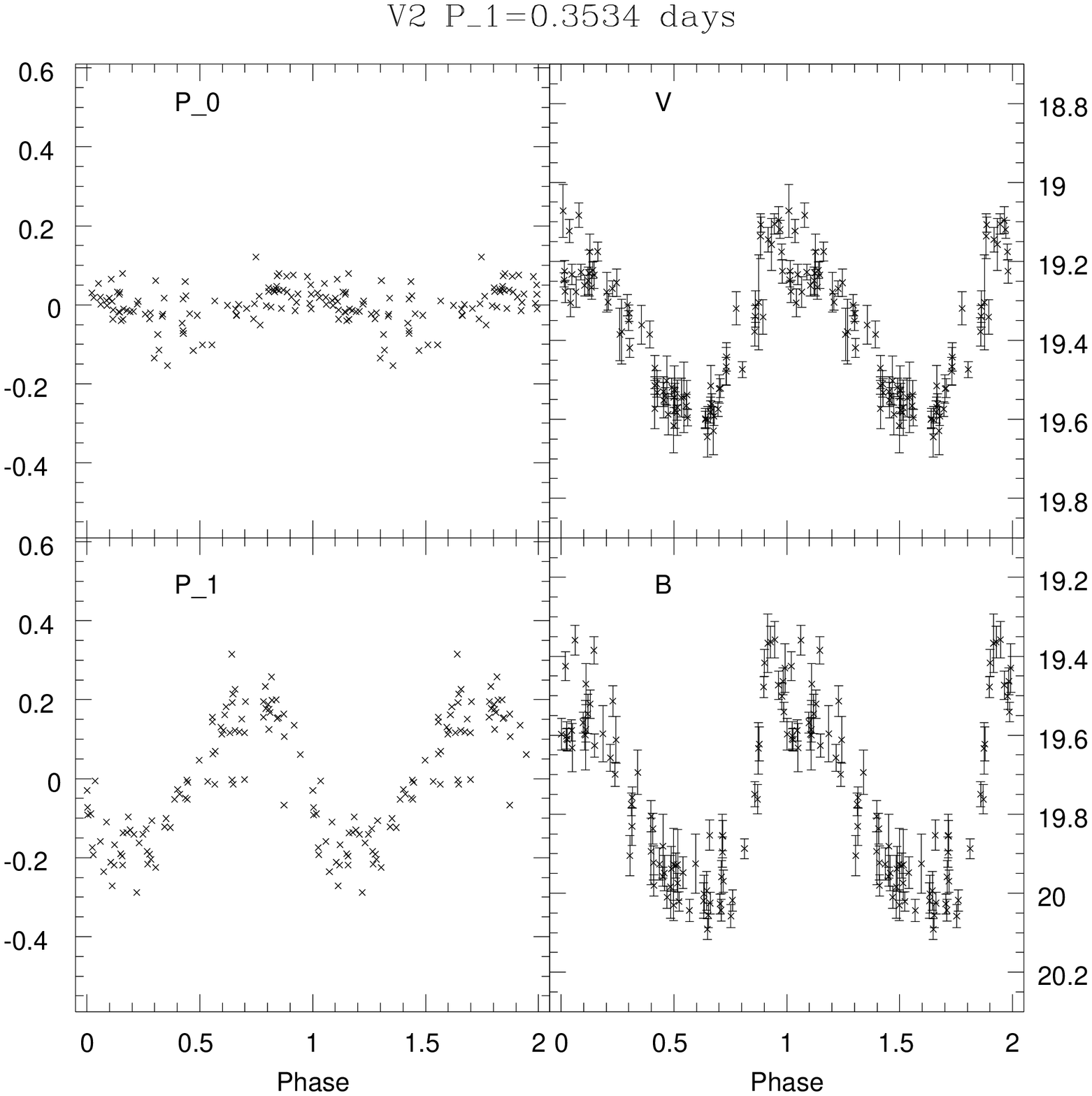}
\includegraphics[width=0.45\textwidth]{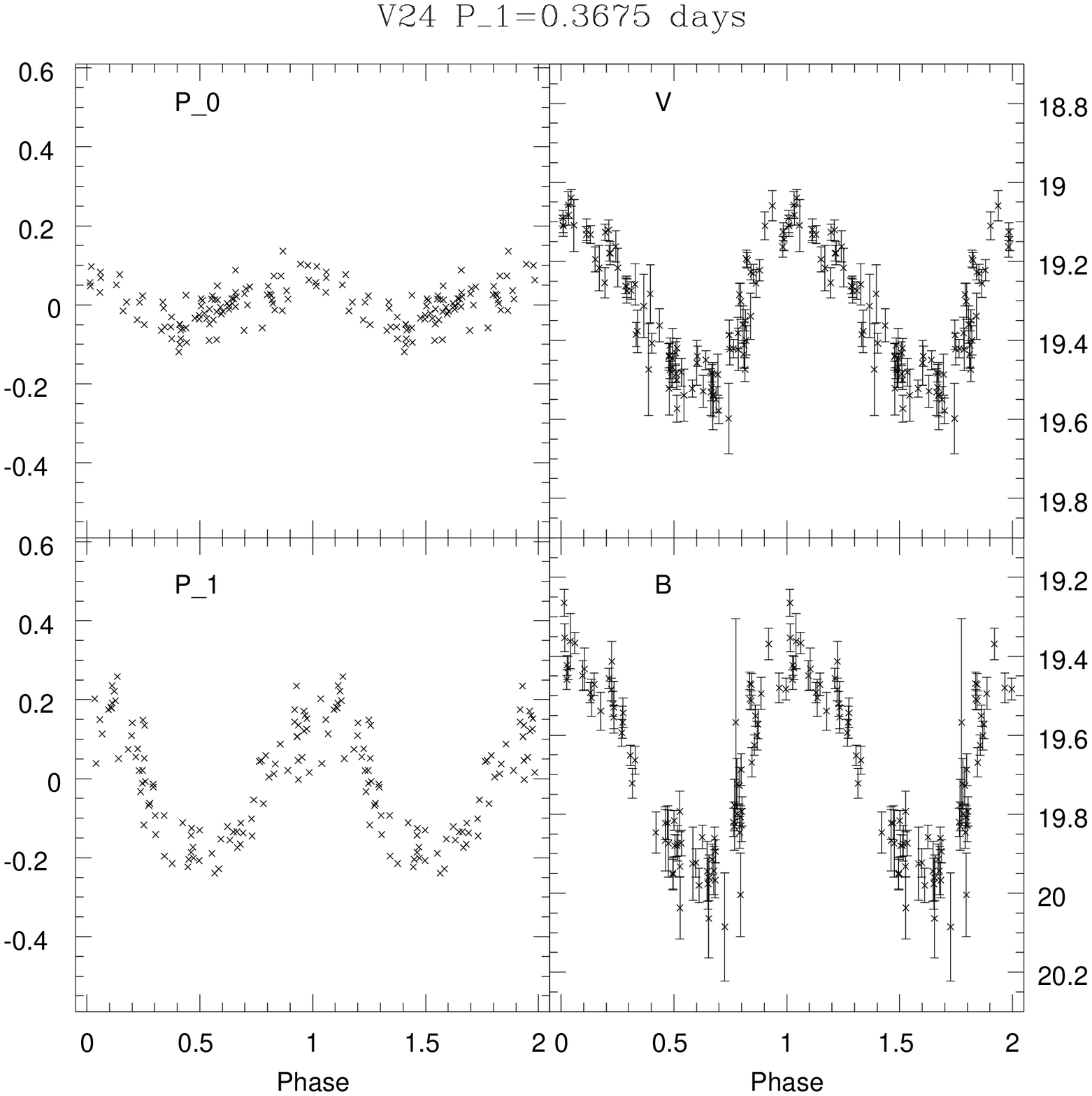}
\caption{Sample light curves for the RRd stars in NGC 1466. The right-hand panels for each star show the raw light curves, as plotted with the first overtone mode period (the first overtone amplitude is larger than the fundamental mode amplitude for all RRd stars in NGC 1466), while the left-hand panels show the deconvolved light curves. (The full set of light curves can be found in the electronic version of this paper.)}
\label{dcurves}
\end{center}
\end{figure*}

\begin{figure*}
\begin{center}
\includegraphics[width=0.45\textwidth]{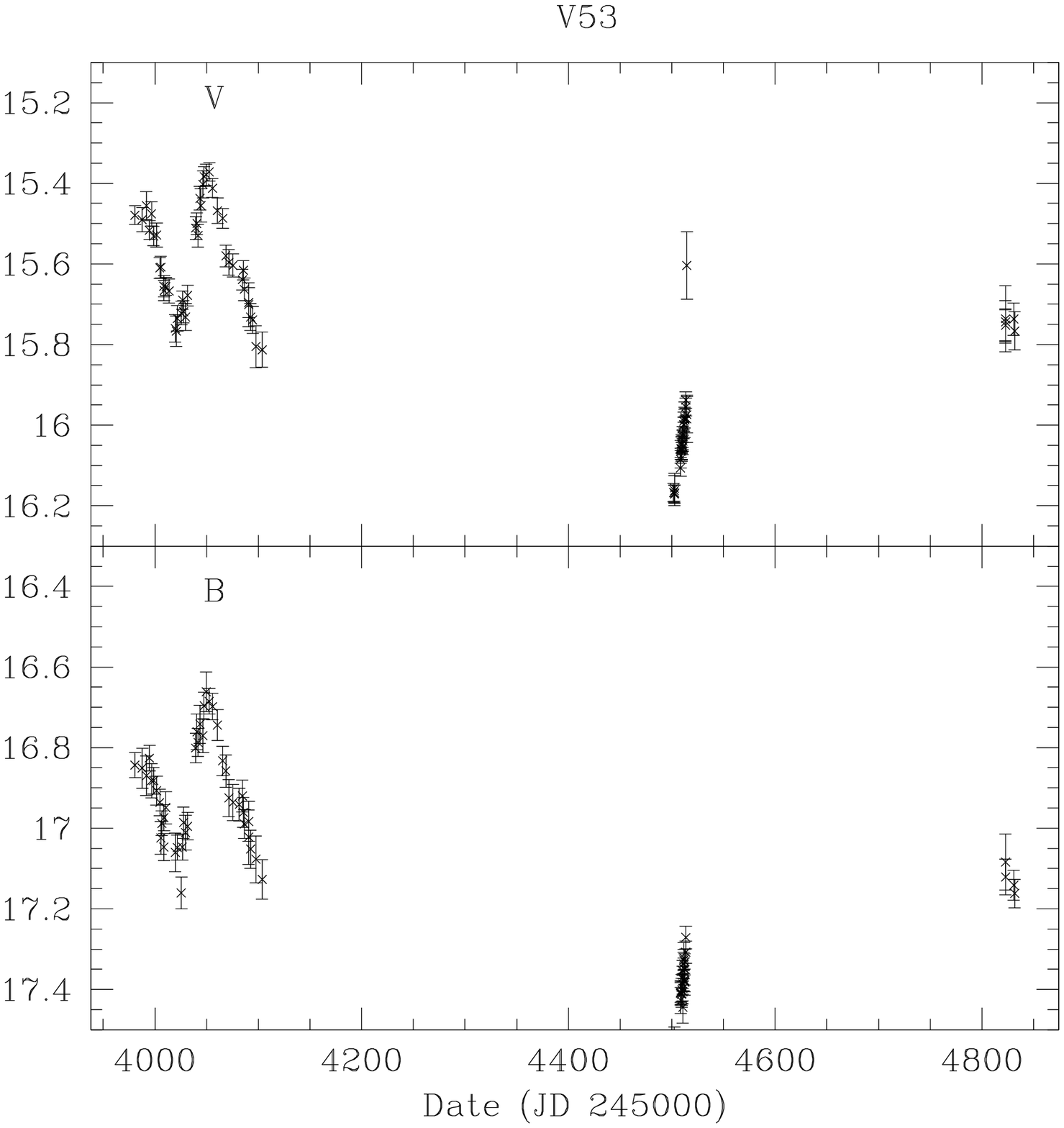}
\includegraphics[width=0.45\textwidth]{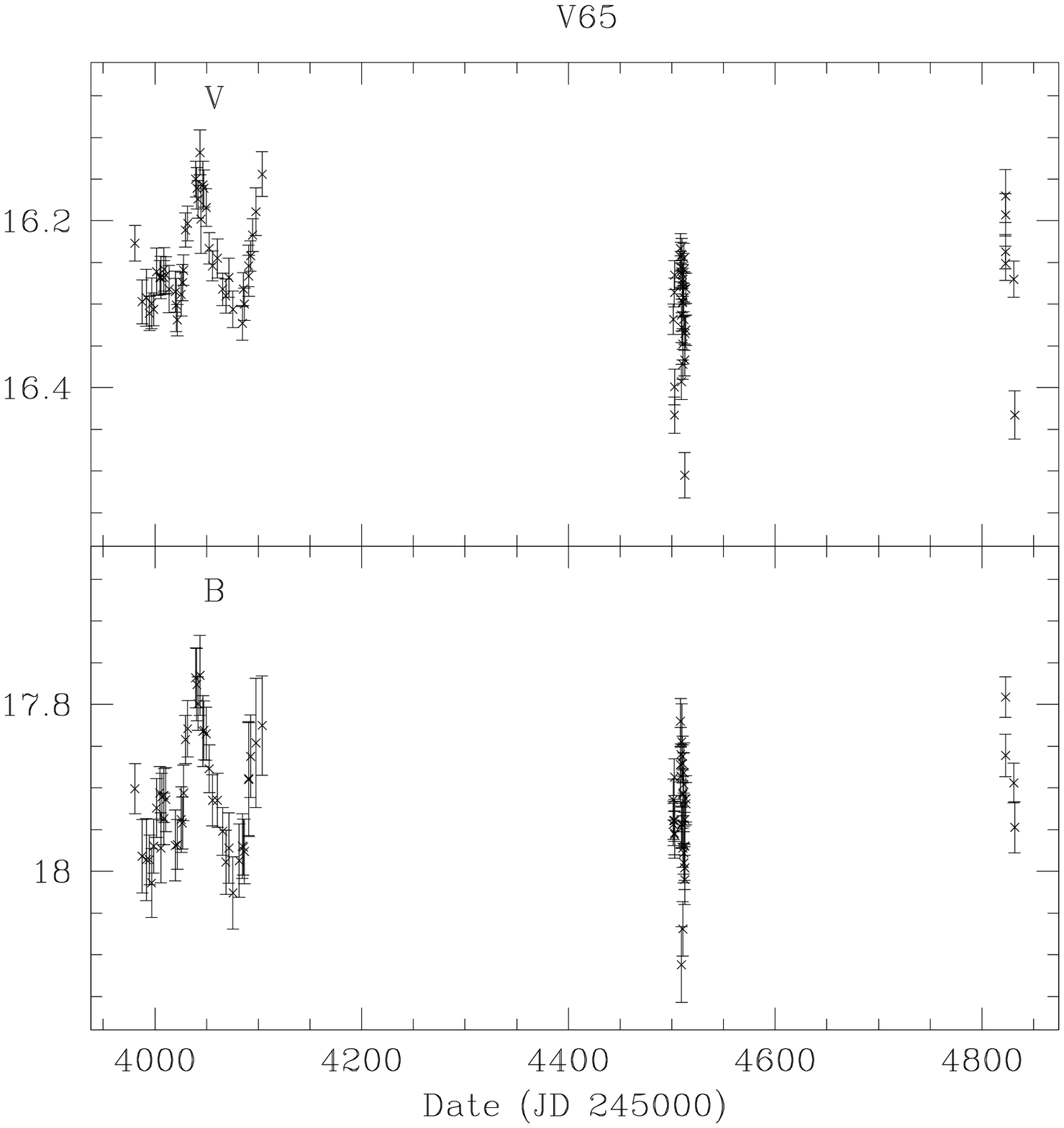}
\includegraphics[width=0.45\textwidth]{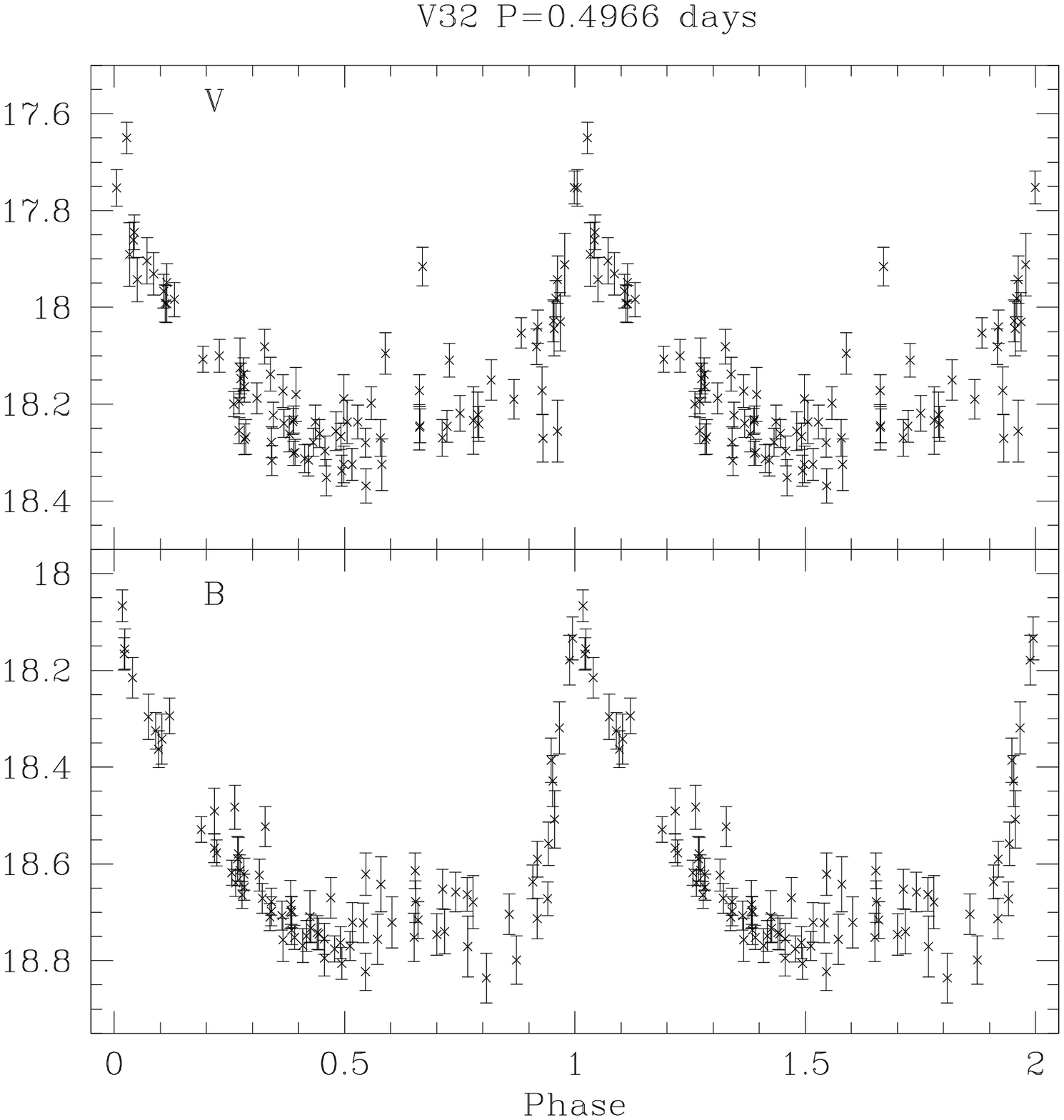}
\includegraphics[width=0.45\textwidth]{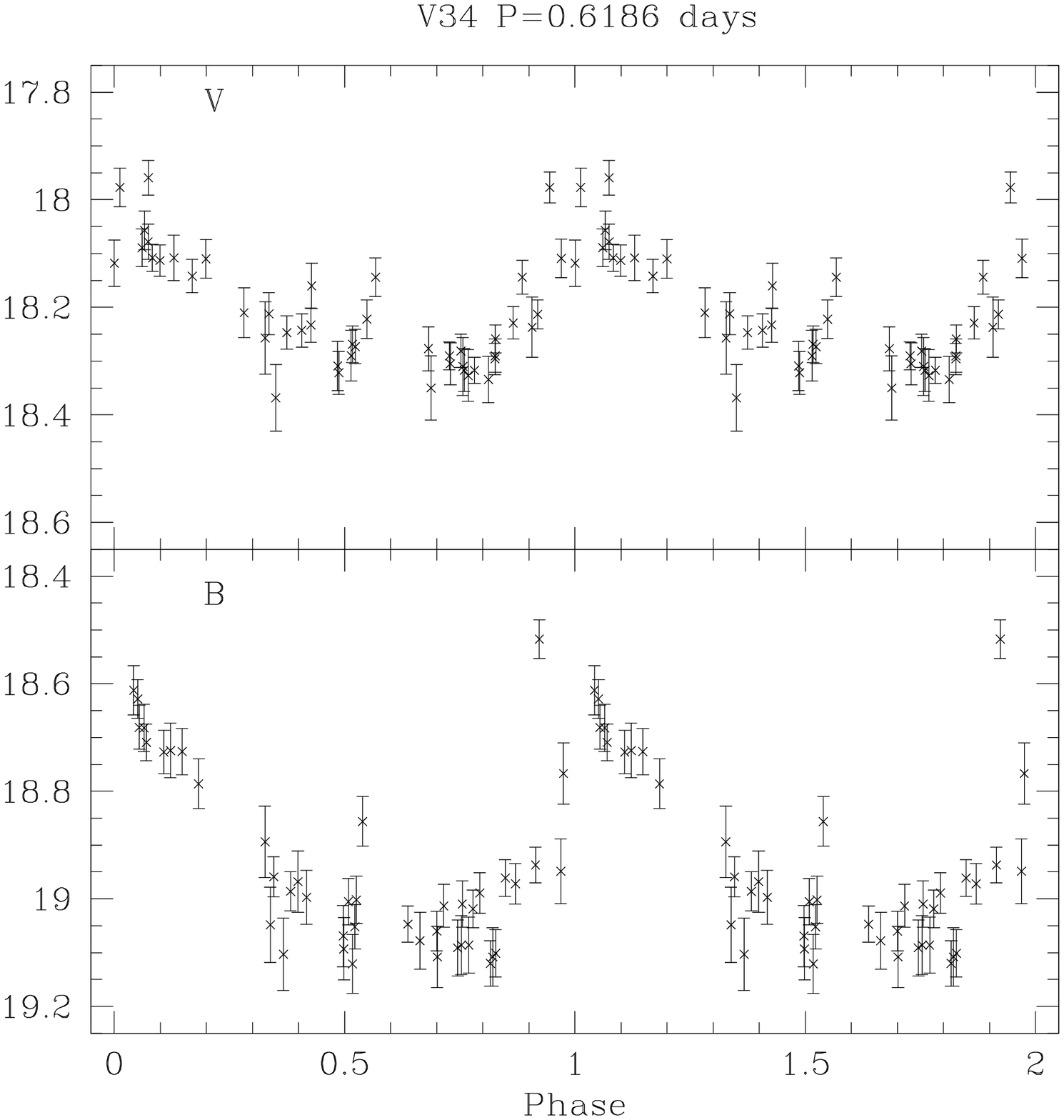}
\includegraphics[width=0.45\textwidth]{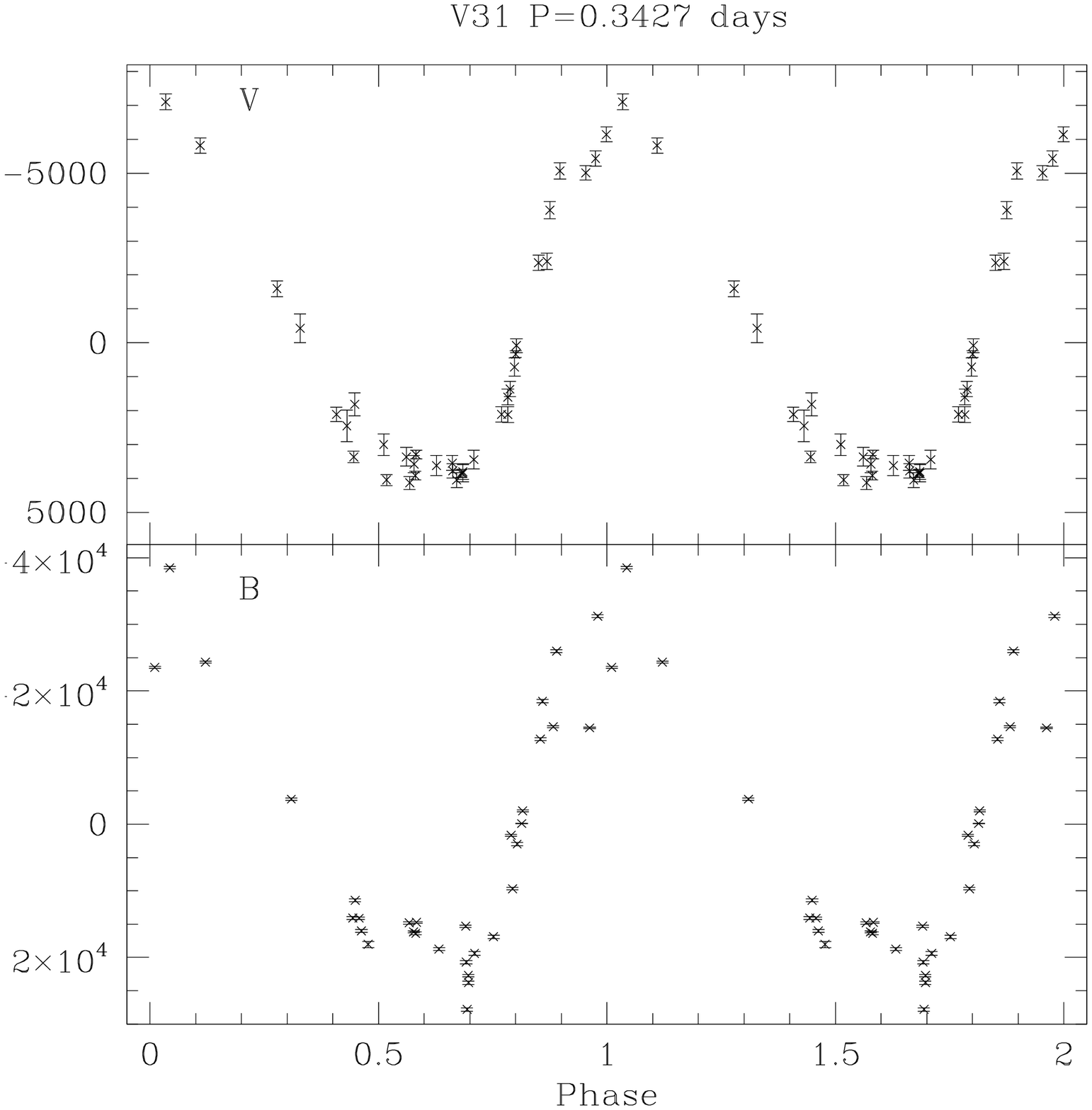}
\includegraphics[width=0.45\textwidth]{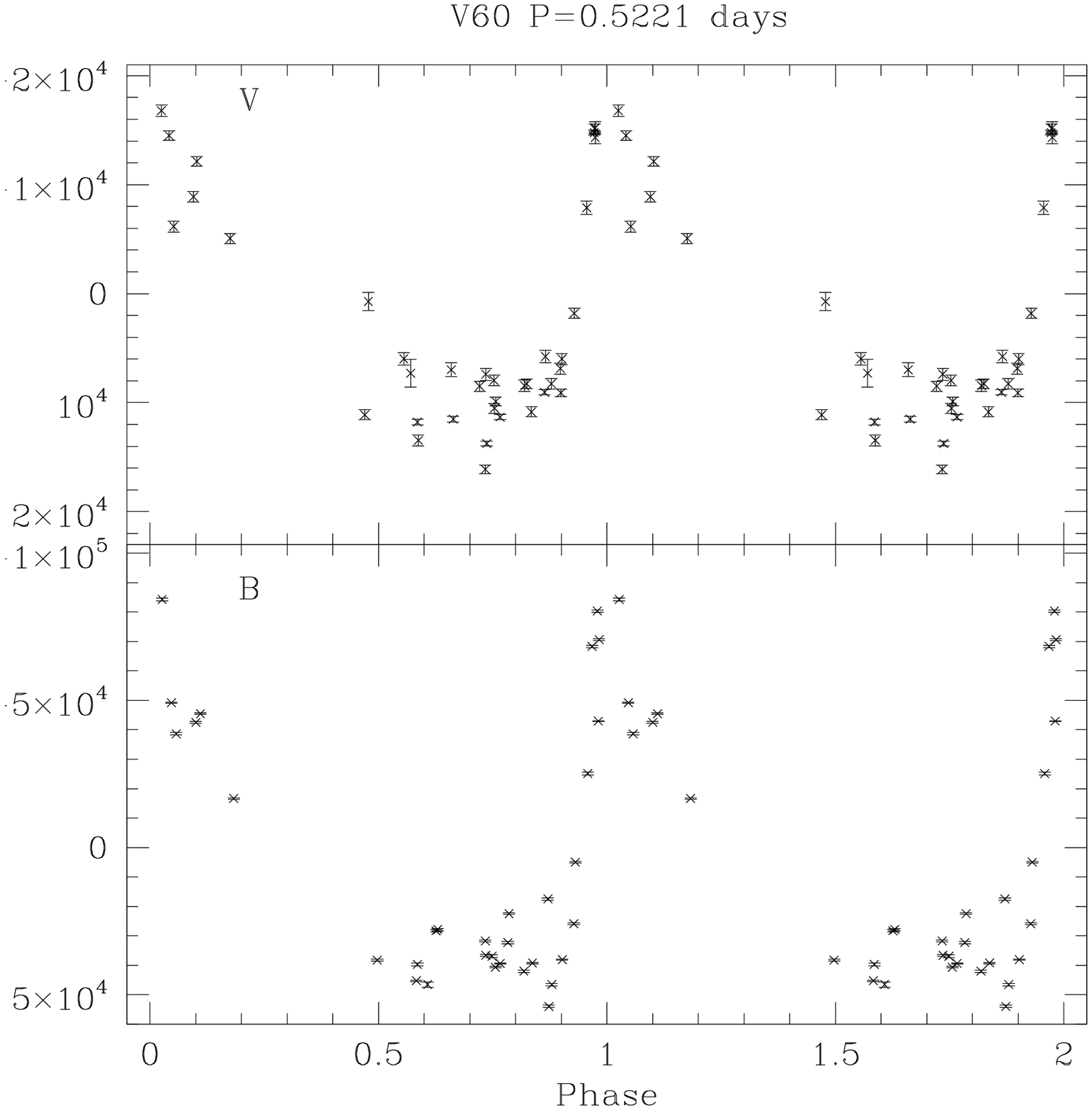}
\caption{Sample light curves for the non-RR Lyrae variable stars in NGC 1466.  The top row shows light curves for our two long-period variables (V$53$ and V$65$).  The other two rows show light curves for the variables of less certain classification, including our candidate anomalous Cepheid (V$32$).  The light curves for V$31$ and V$60$ are from ISIS and thus are plotted in differential fluxes instead of magnitudes.  (The full set of light curves can be found in the electronic version of this paper.)}
\label{othercurves}
\end{center}
\end{figure*}

A total of $62$ variables were found, including $49$ RR Lyraes, $1$ additional candidate RR Lyrae, $2$ long-period variables, a candidate anomalous Cepheid, and $9$ variables of unknown classification.  We were able for the first time to identify double-mode (RRd) variables among the RR Lyrae population of NGC 1466. Of the confirmed RR Lyrae stars, $30$ were RRab stars, $11$ were RRc stars, and $8$ were RRd's.  The RRd variable stars are discussed separately in section \ref{sec:rrd}.  Figures \ref{abcurves}-\ref{othercurves} show the light curves for the RRab , RRc, and the other variable stars.  Table \ref{vartable} lists the identified variable stars, except for the RRd stars, as well as their classification, period, $V$ and $B$ amplitudes,  intensity-weighted $V$ and $B$ mean magnitudes, and magnitude-weighted mean $B-V$ color. The intensity-weighted mean magnitudes and the magnitude-weighted mean color were obtained through the fitting of the light curves with template light curves \citep{ly98}.  (For the relation between these average quantities and the color of the equivalent static star, the reader is referred to Bono et al. 1995.) Notes on some of the individual stars are in the following subsections.  Table \ref{phottable} contains the photometric data for the variable stars.

\begin{deluxetable}{lccccc}
\tablewidth{0pc}
\tabletypesize{\scriptsize}
\tablecaption{Photometry of the Variable Stars}
\tablehead{\colhead{ID} & \colhead{Filter} & \colhead{HJD} & \colhead{Phase} & \colhead{Mag} & \colhead{Mag Error} }
\startdata
V01 & $V$ & $2453980.8494$ & $0.13881$ & $19.252$ & $0.034$ \\
V01 & $V$ & $2453987.7614$ & $0.69800$ & $19.841$ & $0.105$ \\
V01 & $V$ & $2453991.7381$ & $0.92371$ & $19.199$ & $0.052$ \\
V01 & $V$ & $2453994.7980$ & $0.48359$ & $19.552$ & $0.035$ \\
V01 & $V$ & $2453996.7729$ & $0.07200$ & $19.350$ & $0.042$ \\
\enddata
\tablecomments{Maximum light occurs at a phase of $0$.  This table with both the $V$ and $B$-band photometry is published in its entirety in the electronic edition.  The photometry for V$31$, V$51$, V$54$, V$60$, and V$67$ is presented in relative fluxes instead of magnitudes.}
\label{phottable}
\end{deluxetable}

\begin{figure*}
\epsscale{1.0}
\plotone{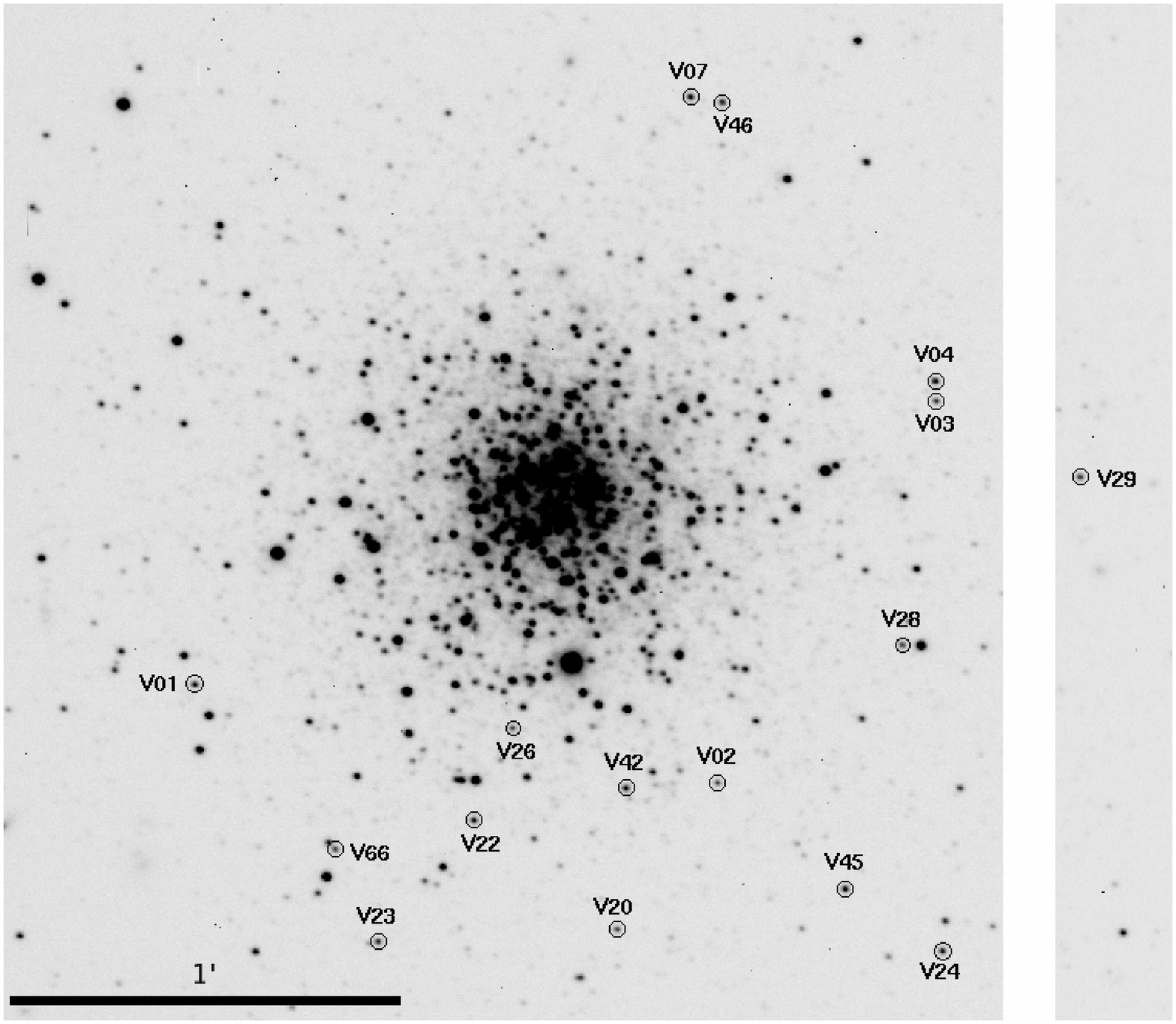}
\caption{Finding chart for the variable stars in the outer regions of NGC 1466.  North is up and East is to the left.  The white gap is due to the finding chart being made from a SOAR image and represents the $7.8$ arcsec mounting gap between the two CCDs of the SOI camera.  The locations of the variables in the inner portion of the cluster are shown in Fig. \ref{findin}.}
\label{findout}
\end{figure*}

\begin{figure*}
\epsscale{1.0}
\plotone{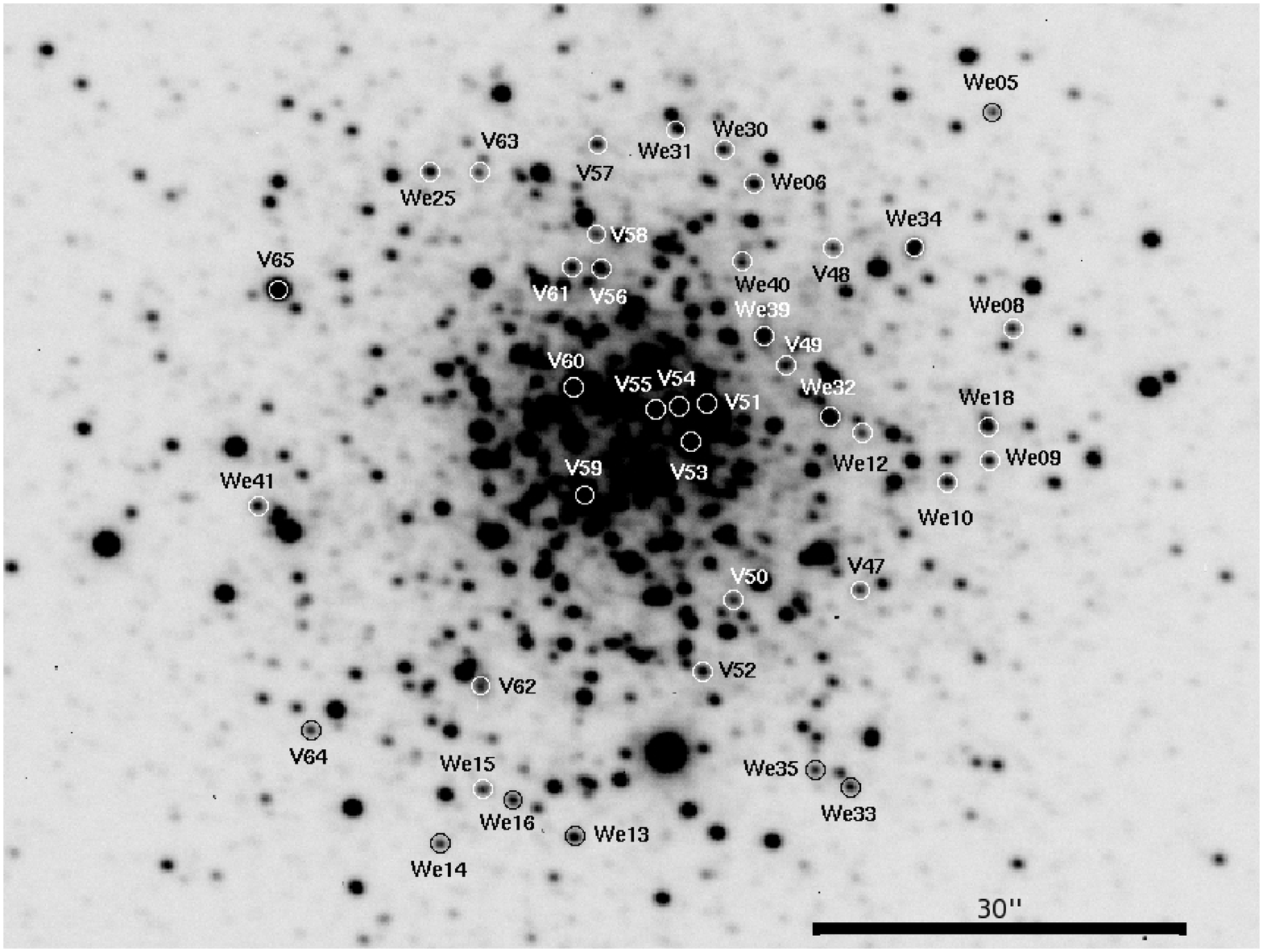}
\caption{Finding chart for the variable stars in the inner portion of NGC 1466.  North is up and East is to the left.}
\label{findin}
\end{figure*}

We use a naming system that is an extension of the one used in \citet{we71}.  Walker identified the variable stars found by Wesselink by their number in that paper, preceded by a ``We''.  However, Walker did not use an extension of Wesselink's numbering scheme for the new variables that he discovered.  Instead, he denoted the new variables by their number in his photometry Table 3.  It is, however, convenient to have an identification system that refers only to the known variable stars in NGC 1466.  Therefore we have extended Wesselink's numbering system, adding the additional variables that are not included in \citet{we71} to the end of Wesselink's orignal list of variables.  All variables are named in the form of Vxx, with stars up through V$42$ being ones that were originally found by Wesselink.  A cross identification of the original Walker numbers and our variable star numbers is included in Table \ref{vartable}.  Figures \ref{findout} and \ref{findin} show the locations of all of the variable stars listed in the table.

\subsection{RR Lyrae Stars}

We found $41$ of Walker's (1992) $42$ RR Lyrae stars along with $6$ new RRab stars, $1$ new RRc, and $2$ new RRd stars. Six of the previously identified RR Lyrae stars in NGC 1466 are reclassified as RRd stars. We did not find Walker's variable $145$, which appears to be a blend of two stars in our images, neither of which display variability.  The additional RR Lyrae stars are all located in the central region of the cluster, which was not searched by Walker due to crowding.  The RRab stars have intensity weighted mean magnitudes of $\langle V\rangle=19.331\pm0.02$ and $\langle B\rangle=19.751\pm0.02$, while the mean magnitudes for the RRc stars are $\langle V\rangle=19.305\pm0.02$ and $\langle B\rangle=19.617\pm0.02$.  Walker found $\langle V\rangle=19.33\pm0.02$ and $\langle B\rangle=19.71\pm0.02$ for the RRab stars while the RRcs had $\langle V\rangle=19.32\pm0.02$ and $\langle B\rangle=19.62\pm0.02$, which are in good agreement with our results.

In general our results agreed with those of Walker for the stars that appeared in both of our lists, with the exceptions noted below.

Walker $110$, V$18$, is listed as an unclassified variable for reasons that are stated in the next subsection.

We were unable to determine periods for Walker's (1992) variables $75$ (V$44$) and $146$ (V$12$) due to not having enough data points for accurate period determination.  These two RRab stars were not detected in all images, but the shape of the partial light curves are entirely consistent with their classification as RRab stars.

V$41$ was classified only as a candidate RR Lyrae star due to aliasing issues which prevented us from determining a period or classification as RRab or RRc.  \citet{wa92b} was unable to measure this star due to its proximity to a brighter star.

No significant differences between our periods and those calculated by \citet{wa92b} were found.  V$33$ was the only star that did not turn out to be a misclassified RRd to display a difference in period greater than $0.01$ days.  Walker found a period of $0.38250$ days while we found a period of $0.27841$ days.  We ran a period search on Walker's light curve for this star and found that there is a secondary signal for a period around $0.27$ days, suggesting that this disagreement in periods is due to aliasing in Walker's light curve.  The limited accuracy of the periods determined by both Walker and ourselves prevents the identification of any evolutionary RR Lyrae period changes.

We did not obtain enough observations to definitively determine if any of the RR Lyrae stars have the Blazhko effect.  However, based on the scatter in their light curves, V$1$, V$4$, V$6$, V$8$, V$14$, V$35$, and V$50$ are potential Blazhko stars.  V$55$ and V$61$ also show scatter in their light curves that could be due to the Blazhko effect, but these two stars are located in the crowded region near the cluster center so the scatter could also be due to contamination.

V$40$ features a light curve that has a fair amount of scatter as well as a shape that is unusually symmetric for an RRab star.  Despite its unusual shape, the light curve was best fit with a period of $0.536$ days and thus is tentatively classified as an RRab.

\subsection{RRd Stars}
\label{sec:rrd}

The increased number of observations (over Wesselink 1971 and Walker 1992) allowed for the identification of eight RRd stars in NGC 1466.  Table \ref{rrdtable} lists the identified RRd stars, their fundamental mode and first overtone periods and amplitudes, and their period ratios. Fits to the two periods were calculated using the {\tt period04} program \citep{le04}.   The secondary periods are typically good to $\pm 0.0001$ or $0.0002$ days while the primarly periods are typically slightly more accurate.  

\begin{deluxetable*}{lccccccccccccccc}
\tablewidth{0pc}
\tabletypesize{\scriptsize}
\tablecaption{Photometric Parameters for the RRd Variables in NGC 1466}
\tablehead{\colhead{ID}& \colhead{RA} & \colhead{DEC} & \colhead{Type}& \colhead{$P_{0}$ (days)}& \colhead{$P_{1}$ (days)}&\colhead{$P_{1}/P_{0}$}& \colhead{$A_{V,0}$}& \colhead{$A_{V,1}$}& \colhead{$A_{B,0}$}& \colhead{$A_{B,1}$}& \colhead{$\langle V\rangle $}& \colhead{$\langle B\rangle $}& \colhead{$(B-V)_{mag}$} & \colhead{Other IDs}& \colhead{Max. Light (HJD)} }
\startdata
V2& 03:44:27.4 & $-$71:40:59.7 & RRd& 0.4751& 0.3534& 0.7451& 0.12& 0.42& 0.15&0.53& 19.375& 19.758& 0.389& Wa124& 2454094.4975\\
V3& 03:44:20.2 & $-$71:40:00.2 & RRd& 0.4717& 0.3505& 0.7431& 0.13& 0.41& 0.15 & 0.53 & 19.368& 19.779& 0.416& Wa66& 2454021.7034\\
V7& 03:44:28.4 & $-$71:39:13.3 & RRd&0.4859 &0.3615 &0.7440 &0.35 &0.43 & 0.42 & 0.52 & 19.326 & 19.642 & 0.310  & Wa135& 2454094.4294\\
V10& 03:44:27.4 & $-$71:40:19.0 & RRd&0.4730 &0.3518 &0.7437 &0.26 &0.48 & 0.30 & 0.52 & 19.425 & 19.797 & 0.370 & & 2454094.5513\\
V24& 03:44:19.9 & $-$71:41:25.5 & RRd& 0.4938 & 0.3675& 0.7442 & 0.12& 0.43& 0.15 & 0.60& 19.292& 19.642& 0.357& Wa65& 2454094.4549\\
V25& 03:44:36.4 & $-$71:39:53.9 & RRd&0.4715 &0.3508 &0.7440 &0.20 &0.44 & 0.40 & 0.65 & 19.360 & 19.705 & 0.350 & Wa323& 2454029.6612\\
V29& 03:44:15.5 & $-$71:40:11.9 & RRd&0.4824 &0.3589 &0.7440 &0.45 &0.58 & 0.36 & 0.38 & 19.428 & 19.750 & 0.320 & Wa53& 2454084.3778\\
V47& 03:44:28.9 & $-$71:40:27.8 & RRd&0.4733 &0.3519 &0.7435 &0.42 &0.58 & 0.48 & 0.60 & 19.410 & 19.760 & 0.350 & & 2454094.6365\\
\enddata
\label{rrdtable}
\end{deluxetable*}

Six of the RR Lyrae identified by \citet{wa92b} turned out to be RRd stars. Five were originally identified as RRc stars and one as an RRab.  V$2$, V$3$, V$24$, and V$25$ all have first overtone periods that agree with the RRc period that was found by Walker.  V$7$ was previously classified as an RRc with a period of $0.34925$ days; however, we obtained a first overtone period of $0.3615$ days for it.  Walker classified V$29$ as an RRab star with a period of $0.56030$ days, which does not correspond with either the fundamental or first overtone period that we found.

Of the two new RRd stars, V$10$ was thought to be a potential variable by \citet{we71}.  \citet{wa92b} did not classify this star as a variable due to its proximity to a blue HB star, which made photometry difficult.

\begin{figure}
\epsscale{1.20}
\plotone{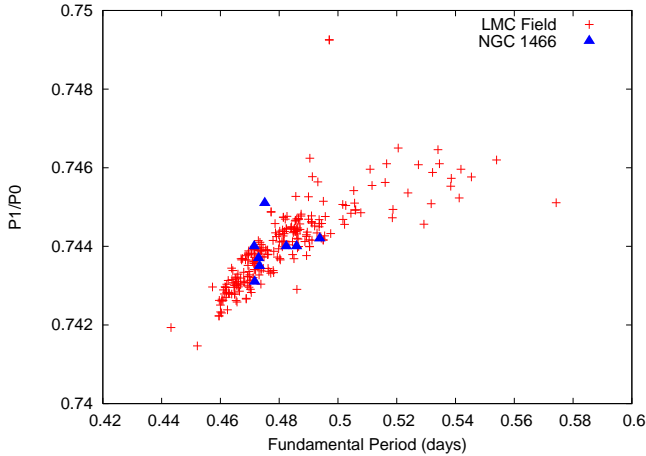}
\caption{Petersen diagram showing the ratio of the first overtone period to the fundamental mode period vs. fundamental mode period for the RRd stars in the cluster (blue triangles).  Also plotted are the RRd stars in the LMC field (red plus symbols) from \citet{so03}.}
\label{peterson}
\end{figure}

Figure \ref{peterson} shows the ratio of the first overtone period ($P_{1}$) to the fundamental mode period ($P_{1}$) vs fundamental period (Petersen diagram) for the RRd stars in NGC 1466;  the RRd stars in the LMC field \citep{so03} are also plotted.  The NGC 1466 RRd stars fall in the same area of the Petersen diagram as the majority of the LMC field stars.  This is the same region where RRd stars in Milky Way Oosterhoff-I clusters tend to fall (see, e.g., Fig. 1 in \citet{po00}, and Fig. 13 in \citet{cl04}).

\subsection{Other Variables}

Two long-period variables, V$53$ and V$65$, were found in our study of NGC 1466.  We were unable to determine periods for either of these stars; in fact, we do not have enough light curve information to say for sure that their variability is periodic.  Both stars are located near the tip of the giant branch on the CMD, as can be seen in Figure \ref{cmd}, and have several possible periods in the range of $60$ to $90$ days.

V$32$ appears to be a potential anomalous Cepheid (AC), having an average $V$ luminosity that is approximately $1.2$ magnitudes brighter in $V$ than the RR Lyrae stars.  The star has a potential period of either $0.331$ days or $0.497$ days.  The $0.497$ day period would be consistent with the period-luminosity diagram for ACs from \citet{pr05}.  Based on Equations $2$ through $5$ in \citet{pr02} and the distance modulus that we derive for NGC 1466, V$32$ would most likely be a first overtone pulsator.  If the $0.331$ day period is correct, the star would be too bright to be an AC, based on the period-luminosity diagram, and would most likely be a foreground RR Lyrae.  The shape of V$32$'s light curve is similar to an RRab star, which would be inconsistent with a $0.331$ day period.  Based on this we feel that the $0.497$ day period is correct and that this star is either an AC or, less likely, a foreground RRab.

Nine variables of unknown classification were found.  $4$ of these stars were only found by ISIS as they were either located in the crowded cluster center (V$31$, V$51$, and V$54$) or near the bright star in the southeast corner of the images (V$67$).  Thus their light curves are in differential fluxes, which does not allow us to obtain positive classifications for them.  While periods and light curve shape can be used to determine potential classifications, without having a magnitude we cannot distinguish between a cluster RR Lyrae, a foreground RR Lyrae, or an AC, for example.  Notes on the other stars with unknown classification follow.

- V$18$ (Walker $110$) is slightly brighter than the horizontal branch but is very red, $B-V = 1.27$.  It has a period of $0.48$ days and an amplitude of $0.15$ mag in $V$.  \citet{wa92b} noted that in his images this star had an elongated shape that suggested a likely unresolved companion.  Our images also show this star to be elongated.  The HST images of \citet{jj99} were checked in order to investigate the possibility of a close companion, however V$18$ did not fall within the HST field.

- V$34$ (Walker $129$) is about one magnitude brighter than the RR Lyrae level and has a $B-V = 0.70$.  \citet{we71} had labeled this star as a variable but \citet{wa92b} considered it a constant RHB star.  The period, $0.61594$ days, and light curve shape of this star suggest that it could be an RRab star whose color and brightness are affected by contamination.

\begin{figure}
\plotone{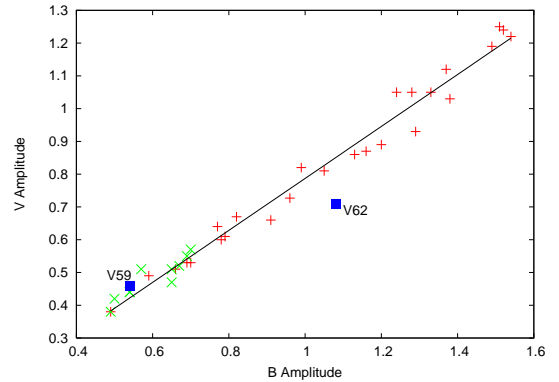}
\caption{$V$ vs $B$ amplitudes for the RRab stars (plus symbols) and RRc stars (crosses).  Also plotted are two of our variables of unknown classification (labeled squares) for purposes of determining if they are RR Lyrae stars that are affected by blending.  V$59$ follows the trend of the RR Lyrae stars, suggesting that it is not blended, while V$62$ departs from this trend, supporting the explanation that it is blended (see discussion in section 3.3).  The plotted line shows the relationship between $V$ and $B$ amplitudes for NGC 2257 \citep{ne09}, an Oo-int globular cluster in the LMC.  The agreement between this line and the RR Lyrae stars in NGC 1466 suggests a similarity between Oo-int globular clusters.}
\label{bvamp}
\end{figure}

- Star V$59$ is located in the cluster center and has an ISIS light curve that suggests several possible periods.  Its Daophot light curves suggest that it has a similar color to the RR Lyrae stars but is about one magnitude brighter; this is possibly due to contamination.  Our images and the HST images of \citet{jj99} confirm the existence of neighbor stars that could be effecting the photometry of V$59$ through blending.  However, V$59$ displays a $B$ to $V$ amplitude ratio that is similar to the other RR Lyraes in the cluster (Fig. \ref{bvamp}).  V$59$ could also potentially be an AC or a Type II Cepheid.

- V$60$ is located in the crowded cluster center and has a clean ISIS light curve that suggests a period of $0.522$ days.  Its light curve from Daophot suggests a luminosity approximately two magnitudes brighter than the RR Lyrae stars but the light curve is messy, possibly due to blending.

- Star V$62$ is located in the cluster center and has a light curve that suggests that it is a variable with a period near $0.5$ days.  The period and the light curve shape suggest that V$62$ is possibly an RR Lyrae but the light curve is messy and it is brighter than the other RR Lyrae stars, likely due to contamination by nearby stars.  The contamination explanation is supported by the fact that V$62$ has a $B$ amplitude that is slightly too large large for its $V$ amplitude, compared to the RR Lyrae stars in NGC 1466.  The HST images of \citet{jj99} confirm a close neighbor for V$62$ which does not appear as a separate star on our images.

\section{Fourier Decomposition}

The RRab light curves were fit with a Fourier series of the form
\begin{equation}
m(t)=A_{0}+\sum_{j=1}^{n}A_{j}\sin (j\omega t+\phi_{j}),
\end{equation}
where $m(t)$ is the magnitude as a function of time and $\omega=2\pi/P$, where $P$ is the pulsation period.  The RRc light curves were fit in a similar fashion but a cosine series was used instead of the sine series.  Fourier decomposition was not carried out on the RRd stars as after we deconvolved their light curves there were not enough data points for accurate Fourier analysis.  The Fourier fitting was done using a Fourier fitting code originally written by Geza Kov\'acs with an analysis wrapper written by Nathan De Lee.  Error analysis for the fits was done based on \citet{pe86}.   A number of fits were performed for each light curve, with a varying number of terms, and the resulting fits were plotted and then checked by eye to determine if it was a successful fit.  Not all light curves were fit successfully, and the optimum number of terms to fit each individual light curve varied.  $13$ RRab and $7$ RRc light curves were successfully fit. Tables \ref{abcoeff} and \ref{ccoeff} give the amplitude ratios, $A_{j1}=A_{j}/A_{1}$, and phase differences, $\phi_{j1}=\phi_{j}-j\phi_{1}$, for the low-order terms ($j=2,3,4$), and the number terms used in the Fourier fit, for the RRab and RRc stars, respectively.  For the RRab stars, we also list the Jurcsik-Kov\'acs $D_m$ value \citep{jk96}.  This value helps to separate RRab stars with ``regular'' light curves from those with ``anomalous'' light curves, with lower $D_m$ values representing more ``regular'' light curves.

\begin{deluxetable*}{lccccccccc}
\tablewidth{0pc}
\tabletypesize{\scriptsize}
\tablecaption{Fourier Coefficients for RRab Variables}
\tablehead{\colhead{ID} & \colhead{$A_{1}$} & \colhead{$A_{21}$} & \colhead{$A_{31}$} & \colhead{$A_{41}$} & \colhead{$\phi_{21}$} & \colhead{$\phi_{31}$} & \colhead{$\phi_{41}$}& \colhead{$D_{max}$}& \colhead{Order}}
\startdata
V4&  0.346& 0.498& 0.301& 0.155& 2.16& 4.77$\pm$0.18& 1.12& 30.46& 8\\
V5&  0.243& 0.380& 0.361& 0.206& 2.31& 5.12$\pm$0.21& 1.69& 49.89& 8\\
V6&  0.422& 0.322& 0.303& 0.174& 2.68& 5.30$\pm$0.27& 1.83& 9.62& 6\\
V8&  0.349& 0.402& 0.324& 0.194& 2.47& 5.02$\pm$0.37& 1.39& 40.96& 7\\
V9&  0.463& 0.416& 0.244& 0.099& 2.30& 4.52$\pm$0.17& 1.10& 7.63& 7\\
V14&  0.355& 0.440& 0.351& 0.205& 2.25& 4.84$\pm$0.14& 1.34& 14.33& 9\\
V19&  0.252& 0.432& 0.177& 0.117& 1.94& 4.93$\pm$0.37& 1.17& 6.53& 7\\
V20&  0.363& 0.502& 0.356& 0.271& 2.26& 4.99$\pm$0.09& 1.26& 45.97& 6\\
V23&  0.424& 0.456& 0.361& 0.237& 2.25& 4.66$\pm$0.08& 1.04& 47.39& 10\\
V26&  0.367& 0.369& 0.249& 0.139& 2.26& 4.80$\pm$0.19& 0.78& 34.99& 7\\
V30&  0.232& 0.457& 0.305& 0.153& 2.36& 5.22$\pm$0.34& 1.93& 23.20& 10\\
V42&  0.204& 0.411& 0.251& 0.098& 2.46& 5.51$\pm$0.16& 2.20& 10.83& 10\\
V56&  0.276& 0.400& 0.282& 0.135& 2.50& 5.38$\pm$0.12& 1.67& 5.31& 8\\
\enddata
\label{abcoeff}
\end{deluxetable*}

\begin{deluxetable*}{lcccccccc}
\tablewidth{0pc}
\tabletypesize{\scriptsize}
\tablecaption{Fourier Coefficients for RRc Variables}
\tablehead{\colhead{ID}&  \colhead{$A_{1}$}& \colhead{$A_{21}$}& \colhead{$A_{31}$}& \colhead{$A_{41}$}& \colhead{$\phi_{21}$}& \colhead{$\phi_{31}$}& \colhead{$\phi_{41}$}& \colhead{Order}}
\startdata
V22&  0.272& 0.179& 0.085& 0.067& 4.67& 2.55$\pm$0.30& 0.89& 6\\
V28&  0.276& 0.252& 0.103& 0.126& 4.73& 3.15$\pm$0.37& 1.55& 6\\
V33&  0.246& 0.255& 0.039& 0.011& 4.38& 2.72$\pm$0.75& 4.95& 8\\
V38&  0.228& 0.114& 0.064& 0.028& 4.58& 4.16$\pm$0.57& 1.74& 6\\
V45&  0.199& 0.090& 0.042& 0.055& 4.41& 2.71$\pm$0.79& 1.53& 6\\
V57&  0.222& 0.139& 0.080& 0.045& 4.71& 3.97$\pm$0.60& 2.01& 7\\
V66&  0.259& 0.181& 0.090& 0.033& 4.30& 3.93$\pm$0.54& 3.24& 9\\
\enddata
\label{ccoeff}
\end{deluxetable*}

\subsection{RRab Variables}
\label{sec:abprop}

\citet{jk96}, \citet{ju98}, and Kov\'acs \& Walker (1999, 2001) provide formulae, determined empirically, for calculating the metallicity, absolute magnitude, and effective temperature of RRab stars using the Fourier coefficients from their $V$ light curves.  [Fe/H], $M_{V}$, $V-K$, and $\log T_{\rm eff}^{\langle V-K\rangle}$ are determined using the following equations, which come from equations (1),(2),(5), and (11) in \citet{ju98}, respectively:
\begin{equation}
{\rm [Fe/H]} = -5.038-5.394\,P+1.345\,\phi_{31} ,
\end{equation}
\begin{equation}
M_{V} = 1.221-1.396\,P-0.477\,A_{1}+0.103\,\phi_{31} ,
\end{equation}
\begin{multline}
(V-K)_{0} = 1.585+1.257\,P-0.273\,A_{1}\\-0.234\,\phi_{31}+0.062\,\phi_{41} ,
\end{multline}
and
\begin{equation}
\log T_{\rm eff}^{\langle V-K\rangle} = 3.9291-0.1112\,(V-K)_{0}-0.0032\,{\rm [Fe/H]}.
\end{equation}
The values for [Fe/H] derived through this method are in the scale of \citet{ju95}, which can be transformed to the more common \citet{zw84} scale by the relation ${\rm [Fe/H]_{J95}} =1.431{\rm [Fe/H]_{ZW84}}+0.880$ \citep{ju95}.  The dereddened $B-V$ and $V-I$ color indices were calculated using equations (6) and (9) from \citet{kw01}:
\begin{equation}
(B-V)_{0}=0.189\,\log P-0.313\,A_{1}+0.293\,A_{3}+0.460
\end{equation}
and
\begin{equation}
(V-I)_{0}=0.253\,\log P-0.388\,A_{1}+0.364\,A_{3}+0.648.
\end{equation}
These color indices are also used to calculate effective temperatures via equations (11) and (12) in \citet{kw01},
\begin{multline}
\log T_{\rm eff}^{\langle B-V\rangle}=3.8840-0.3219\,(B-V)_{0}\\+0.0167\,\log (g)+0.0070\,[M/H]
\end{multline}
and
\begin{multline}
\log T_{\rm eff}^{\langle V-I\rangle}=3.9020-0.2451\,(V-I)_{0}\\+0.0099\,\log (g)-0.0012\,[M/H] ,
\end{multline}
where $g$ is obtained from equation (12) in \citet{kw99},
\begin{multline}
\log (g)=2.9383+0.2297\,\log (M/\msun)\\-0.1098\,\log T_{eff}-1.2185\,\log P ,
\end{multline}
assuming a mass of $M=0.7\,\msun $.  The resulting physical properties for the RRab stars are given in Table \ref{abphysical}.

The mean metallicity of the RRab stars is ${\rm [Fe/H]_{J95}}=-1.40\pm0.09$ which in the Zinn \& West scale is ${\rm [Fe/H]_{ZW84}}=-1.59\pm0.06.$  Previously published metallicities for NGC 1466 showed a wide range of values.  \citet{ol91} reported a metallicity of ${\rm [Fe/H]}=-2.17$ based on spectroscopic observations of the calcium triplet in two giant stars in the cluster whose individual metallicities were ${\rm [Fe/H]}=-1.85$ and $-2.48$.  \citet{wa92b} argued that the more metal-poor of these two stars was not a cluster member based on photometric results.  Walker instead reported that the cluster metallicity was ${\rm [Fe/H]}=-1.85$, based on the remaining star from Olszewski et al. and comparisons of the RR Lyrae stars and the cluster CMD to that of other clusters.  \citet{wo07} used stellar popular synthesis models to find metallicities of ${\rm [Fe/H]}=-1.30$ and $-2.00$, depending on whether they used the full spectrum or just the CN spectrum.  A more metal poor result of ${\rm [Fe/H]}=-2.25$ was adopted by \citet{mg03} for their work with cluster surface brightness profiles.  A far more metal-rich value of ${\rm [Fe/H]}=-1.64$ was found by \citet{sp04} using the equivalent widths of metal lines in integrated spectra.  Our value of ${\rm [Fe/H]}=-1.59$ is most consistent with the Santos \& Piatti result.

\subsection{RRc Variables}

\citet{sc93} demonstrated that the Fourier parameters of the $V$-band light curves of RRc stars could be used to calculate various physical properties for these stars.  Unlike the empirically determined relationships between physical properties and Fourier parameters for RRab stars, the relationships for RRc stars were determined using light curves that were created with hydrodynamic pulsation models.  The mass, luminosity, temperature, and helium parameter of RRc stars can be calculated using the equations:
\begin{multline}
\log T_{\rm eff}=3.265-0.3026\,\log P\\-0.1777\,\log M+0.2402\,\log L ,
\end{multline}
\begin{multline}
\log y=-20.26+4.935\,\log T_{eff}\\-0.2638\,\log M+0.3318\,\log L ,
\end{multline}
\begin{equation}
\log M=0.52\,\log P-0.11\,\phi_{31}+0.39 ,
\end{equation}
and
\begin{equation}
\log L=1.04\,\log P-0.058\,\phi_{31}+2.41 ,
\end{equation}
which are equations (2),(3),(6), and (7) in \citet{sc93}, respectively.  It should be noted that the helium abundance parameter, $y$, is not equal to the helium abundance, $Y$.  Equation (10) in \citet{ko98},
\begin{equation}
M_{V}=1.261-0.961\,P-0.044\,\phi_{21}-4.447\,A_{4} ,
\end{equation}
allows us to calculate the absolute magnitude for the RRc stars.  The metallicity in the \citet{zw84} scale can be calculated using equation (3) from \citet{mw07}:
\begin{multline}
{\rm [Fe/H]_{ZW84}}=52.466\,P^{2}-30.075\,P+0.131\,\phi_{31}^{2}\\+0.982\,\phi_{31}-4.198\,\phi_{31}P+2.424.
\end{multline}
It should be noted that while these equations are commonly used in the literature, the equations for mass, luminosity, and temperature violate the pulsation equation \citep{ca04,ds10}.  While the values calculated for these properties cannot all be physically correct, they can be used for the purpose of making comparisons with the physical properties calculated the same way for RRc stars in other clusters.  The calculated physical properties for the RRc stars are given in Table \ref{cphysical}.

\begin{deluxetable*}{lcccccccc}
\tablewidth{0pc}
\tabletypesize{\scriptsize}
\tablecaption{Derived Physical Properties for RRab Variables}
\tablehead{\colhead{ID}& \colhead{${\rm [Fe/H]_{J95}}$}& \colhead{$\langle M_{V}\rangle$}& \colhead{$\langle V-K\rangle$}& \colhead{$\log T_{\rm eff}^{\langle V-K\rangle}$}& \colhead{$\langle B-V\rangle$}& \colhead{$\log T_{\rm eff}^{\langle B-V\rangle}$}& \colhead{$\langle V-I\rangle$}& \colhead{$\log T_{\rm eff}^{\langle V-I\rangle}$}}
\startdata
V4& -1.717& 0.748& 1.164& 3.805& 0.336& 3.810& 0.490& 3.811\\
V5& -1.278& 0.824& 1.153& 3.805& 0.365& 3.804& 0.526& 3.802\\
V6& -0.764& 0.828& 1.008& 3.819& 0.313& 3.825& 0.461& 3.818\\
V8& -1.085& 0.846& 1.054& 3.815& 0.330& 3.817& 0.482& 3.813\\
V9& -1.774& 0.736& 1.125& 3.810& 0.295& 3.824& 0.438& 3.825\\
V14& -1.577& 0.762& 1.149& 3.806& 0.338& 3.811& 0.493& 3.811\\
V19& -1.725& 0.749& 1.209& 3.800& 0.354& 3.804& 0.513& 3.805\\
V20& -1.511& 0.737& 1.139& 3.807& 0.341& 3.810& 0.496& 3.810\\
V23& -1.434& 0.810& 1.064& 3.815& 0.314& 3.821& 0.462& 3.819\\
V26& -1.359& 0.823& 1.057& 3.816& 0.317& 3.820& 0.466& 3.818\\
V30& -1.759& 0.679& 1.292& 3.791& 0.378& 3.795& 0.544& 3.797\\
V42& -1.056& 0.805& 1.175& 3.802& 0.374& 3.802& 0.538& 3.799\\
V56& -1.155& 0.775& 1.136& 3.806& 0.357& 3.807& 0.517& 3.804\\
\hline
Mean& -1.400$\pm$0.088& 0.779$\pm$0.013& 1.133$\pm$0.021& 3.808$\pm$0.002& 0.339$\pm$0.007& 3.812$\pm$0.003& 0.494$\pm$0.009& 3.810$\pm$0.002\\
\enddata
\label{abphysical}
\end{deluxetable*}

\begin{deluxetable*}{lcccccc}
\tablecaption{Derived Physical Properties for RRc Variables}
\tablewidth{0pc}
\tabletypesize{\scriptsize}
\tablehead{\colhead{ID}& \colhead{${\rm [Fe/H]_{ZW84}}$}& \colhead{$\langle M_{V}\rangle$} & \colhead{$M/\msun$}& \colhead{$\log(L/\lsun)$}& \colhead{$\log T_{\rm eff}$}& \colhead{Y}}
\startdata
V22& -1.970& 0.657& 0.724& 1.762& 3.859& 0.254\\
V28& -1.669& 0.591& 0.611& 1.713& 3.864& 0.272\\
V33& -1.420& 0.788& 0.633& 1.675& 3.871& 0.281\\
V38& -1.419& 0.696& 0.495& 1.693& 3.864& 0.284\\
V45& -1.764& 0.720& 0.672& 1.724& 3.864& 0.266\\
V57& -1.614& 0.666& 0.526& 1.715& 3.862& 0.276\\
V66& -1.425& 0.711& 0.515& 1.690& 3.865& 0.283\\
\hline
\rm{Mean}& -1.612$\pm$0.079& 0.690$\pm$0.023& 0.597$\pm$0.033& 1.710$\pm$0.011& 3.86$\pm$0.001& 0.274$\pm$0.004\\
\enddata
\label{cphysical}
\end{deluxetable*}

It has been noted in the literature that RRc masses calculated using the \citet{sc93} equations can at time be too small,$\simeq 0.5\msun$, approaching the mass of a degenerate helium core at the helium flash \citep{co03}.  Table \ref{cphysical} shows that this occurs for three of our stars.  This is likely due to a problem with the equations used but still produces values that are suitable for comparison purposes.

The mean metallicity of the RRc stars is ${\rm [Fe/H]_{ZW84}}=-1.61\pm0.08$, which is in good agreement with the value obtained for the RRab stars.

\section{Distance Modulus}

From the Fourier fits we obtain a mean $V$ absolute magnitude of $\langle M_{V}\rangle=0.690\pm0.023$ for the RRc stars and  $\langle M_{V}\rangle=0.779\pm0.013$ for the RRab stars.  The zero points used for these calibrations are different and there is some debate as to which one, if either, is correct.  Instead we use the calibration of the absolute magnitude-metallicity relation for RR Lyrae stars by \citet{cc08}.  The average of our Fourier derived metallicities for the RRab and RRc stars is ${\rm [Fe/H]_{ZW84}}=-1.60\pm0.05$, which gives us an absolute magnitude of $M_{V}=0.62\pm0.14$. The average apparent magnitude of the RRab and RRc stars are $\langle V\rangle=19.331\pm0.02$ and $\langle V\rangle=19.305\pm0.02$, respectively.  Since the average magnitude for the RRab and RRc stars are essentially the same, we combine them for the purpose of calculating the distance, resulting in an average magnitude of $V=19.324\pm0.013$.  Using the reddening value of $E(B-V)=0.09\pm0.02$ from \citet{wa92b} and a standard extinction law with $A_{V}/E(B-V)=3.1$, we obtain a reddening-corrected distance modulus of $(m-M)_{0}=18.43\pm0.15$.  This distance modulus is approximately equal to the distance modulus of $(m-M)_{LMC}=18.44\pm0.11$ that \citet{cc08} derived for the LMC.





\section{Oosterhoff Classification}
The average periods for the RR Lyrae stars in NGC~ 1466 are $\langle P_{ab}\rangle = 0.591$ days and $\langle P_{c}\rangle = 0.335$ days.  We found $30$ RRab, $11$ RRc, and $8$ RRd stars, giving the cluster a $N_{c+d}/N_{c+d+ab} = 0.39$.  All three of these values are consistent with an Oosterhoff-intermediate (Oo-int) classification for NGC 1466.  The minimum period for RRab stars has also been shown to be a good indicator of Oosterhoff class.  NGC 1466 has a minimum period for RRab stars of $P_{ab,min}=0.4934$ days, which is consistent with it being an Oo-int object (Catelan et al. 2011, in preparation).

\begin{figure}
\includegraphics[width=0.45\textwidth]{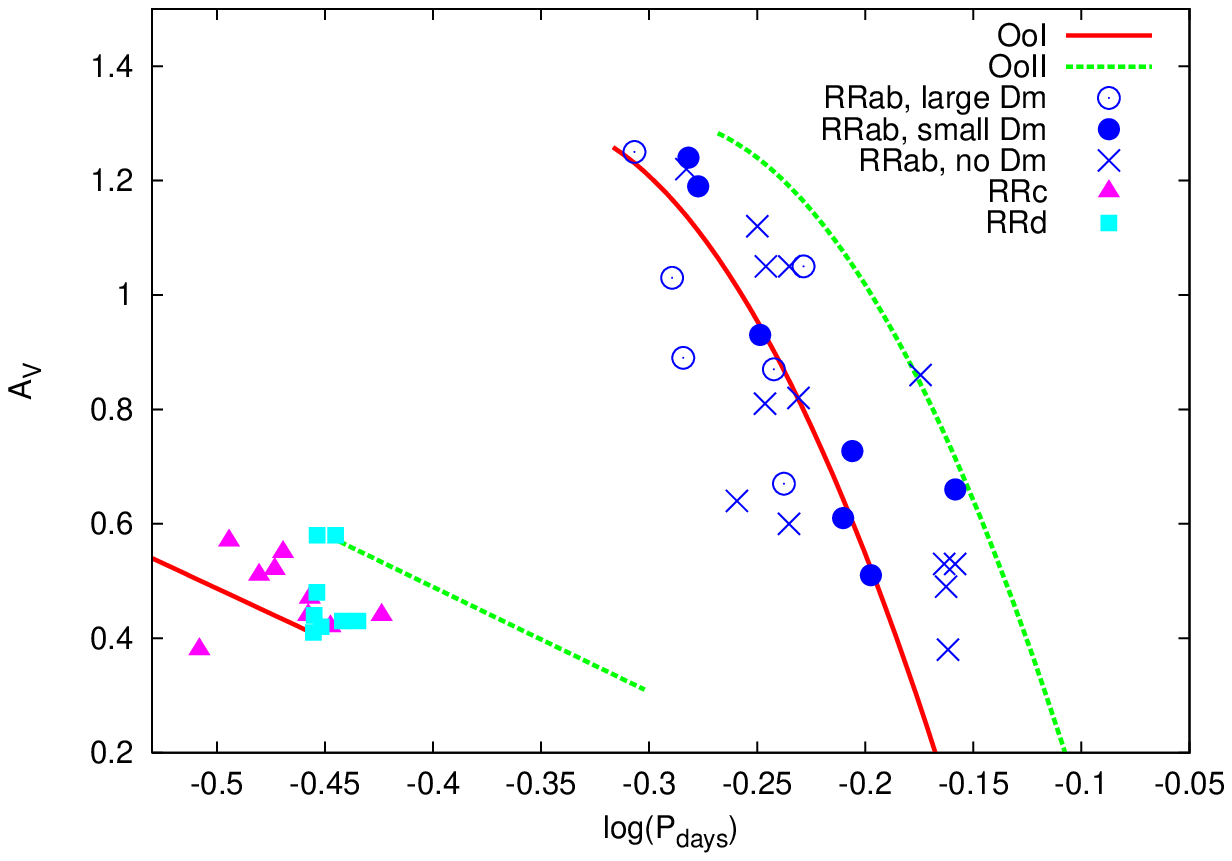}
\includegraphics[width=0.45\textwidth]{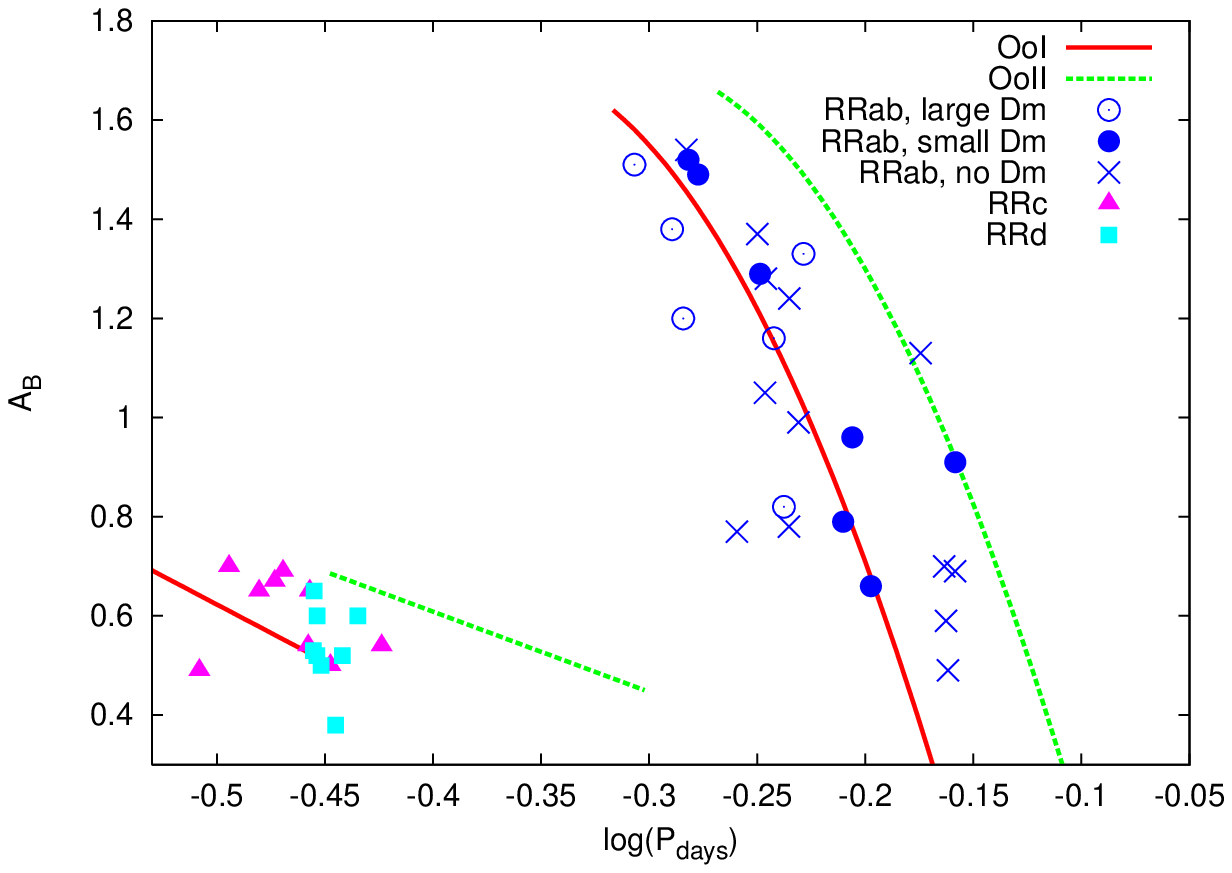}
\caption{Bailey diagram, log period vs $V$-band amplitude (top panel), for the RR Lyrae stars in NGC 1466.  Solid red lines and dashed green lines indicate the typical position for RR Lyrae stars in Oosterhoff I and Oosterhoff II clusters, respectively \citep{cc05,zo10}.  RRab stars for which we obtained a value of $D_{m}<25$ are indicated with filled circles, those with a $D_{m}>25$ are indicated with open circles, and the RRab stars for which we were unable to obtain a good Fourier fit are indicated by crosses.  The bottom panel is the same as the top panel but using the $B$-band amplitudes.}
\label{vperamp}
\end{figure}

Another indicator of Oosterhoff status is the position of the RR Lyrae stars in the Bailey period-amplitude diagram.  Figure \ref{vperamp} shows the diagrams for NGC 1466 using both the $V$ and $B$-band amplitudes.  The typical location of the RRab and RRc stars in Oosterhoff I (Oo-I) and II (Oo-II) clusters are indicated by solid red lines and dashed green lines, respectively \citep{cc05,zo10}.  In both diagrams the RRab stars display a wide scatter.  Many are located between the Oo-I and O-II loci, which is consistent with an Oo-int object.  However, there is a significant number of RRab stars that are clustered around the Oo-I line, which is more typical for an Oo-I object.  The RRc stars fall almost exclusively between the reference lines, as is expected for an Oo-int object.

We found an HB type of $(B-R)/(B+V+R)=0.38\pm0.05$ for NGC 1466.  This was determined using the stars in an annulus with an inner radius of $10$ arcsec and an outer radius of $75$ arcsec; this represented a region for which we felt our sample of the HB stars was complete.   This HB type, combined with the metallicity of ${\rm [Fe/H]}\approx-1.60$ that we obtained from the RR Lyrae stars, places the cluster outside of the ``forbidden region'' where Oo-int objects tend to fall in the HB type-metallicity diagram (Fig. 7 in Catelan 2009).  If one used a lower metallicity similar to what was favored by \citet{wa92b} or \citet{ol91}, NGC 1466 would be located much closer to the forbidden region.  A lower metallicity would also be consistent with the theoretical predictions by \citet{bo94} for Oo-int behavior, whereas the higher metallicity would indicate an HB morphology that is too red with respect to the predictions of the models (see Fig. 8 in Catelan 2009).

\begin{figure}
\epsscale{1.15}
\plottwo{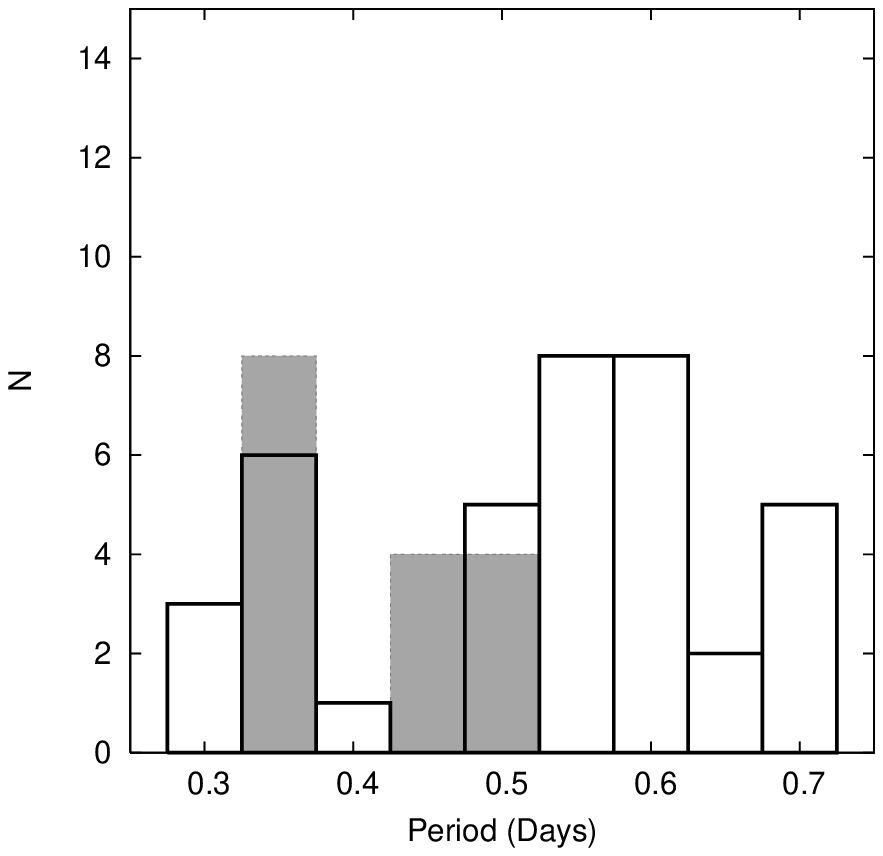}{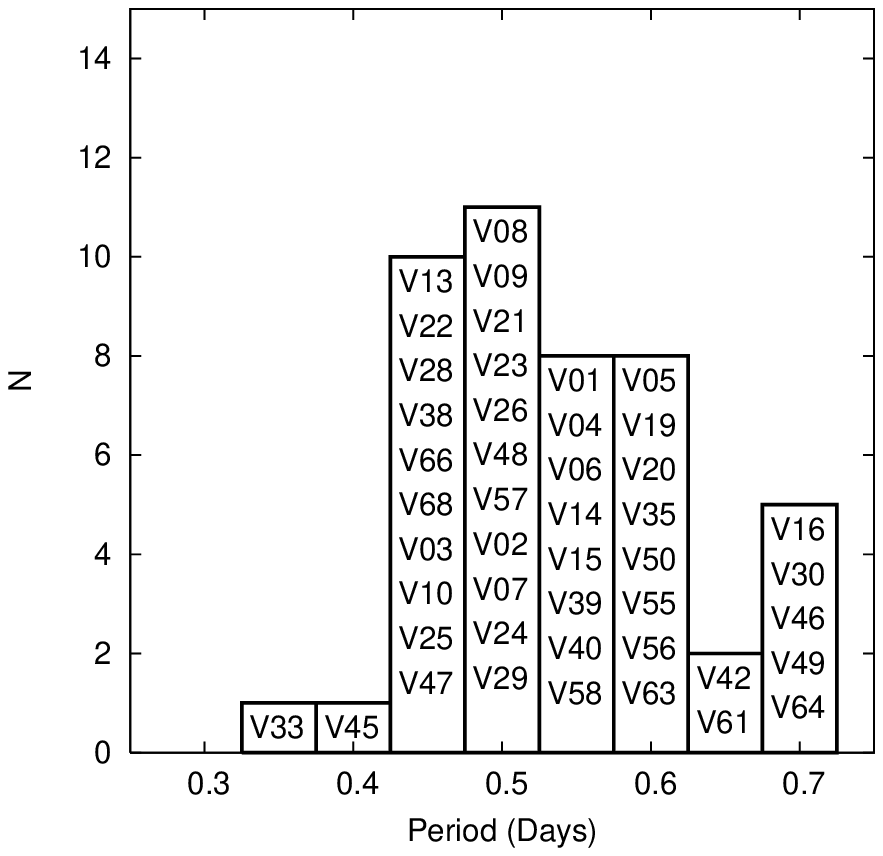}
\caption{Distribution of raw periods (left) and fundamentalized periods (right) of the RR Lyrae stars found in NGC 1466.  The empty boxes in the left panel show the distribution for the RRab and RRc stars while the shaded boxes represent the RRd stars; both the fundamental and first overtone periods are plotted for the RRd stars.  All of the RR Lyrae stars are included in the right panel with the variable stars that are included in each bin labeled.}
\label{hist}
\end{figure}

Figure \ref{hist} shows the distribution of both the raw and the fundamentalized periods for the RR Lyrae stars in NGC 1466.  A sharply peaked fundamentalized period distribution is not present, in contrast to what was found in \citet{mc04} in a number of OoI and OoII clusters. The distribution of the raw periods for the RRab stars displays two peaks, a broad one around $0.55$ days and a sharper peak at $0.70$ days.  This is similar to the distribution seen in NGC 1835 \citep{so03} although the shorter period peak for NGC 1466 is broader and its longer period peak occurs at a slightly longer period.  Soszy\'{n}ski suggested that this two-peaked distribution may be an indicator of a sum of Oo-I and Oo-II characteristics, but our Bailey diagram shows no evidence of an Oo-II population.

Tables 7 and 8 in \citet{co10} show the mean physical parameters for RRc and RRab stars, respectively, in a selection of previously studied globular clusters.  The parameters that we calculated for NGC 1466 fall between those for Oo-I and Oo-II clusters, consistent with NGC 1466 having an Oo-int classification.  A more thorough comparison of the physical parameters for the RR Lyrae stars in NGC 1466 and the other globular clusters in our study will be the subject of a future paper in this series.

\section{Conclusion}

A photometric study of the LMC globular cluster NGC 1466 was conducted in order to locate variable stars, with a specific interest in the cluster's RR Lyrae population.  $49$ RR Lyrae stars and $1$ candidate RR Lyrae were found.  We discovered $9$ RR Lyrae stars which were not found by \citet{wa92b}, previously the most extensive search for RR Lyrae in NGC 1466.  Of our $49$ RR Lyrae stars, $8$ are RRd stars, making this the first time double-mode RR Lyrae stars were found in this cluster.

A candidate anomalous Cepheid variable was found in NGC 1466.  ACs are rare in globular clusters, having previously been found only in NGC 1786, NGC 5466, and $\omega$ Cen \citep{ne94,cc01}.  Interestingly, these three systems are OoII, though this of course may be just a coincidence, particularly since it is believed that ACs in globular clusters should be the progeny of blue straggler stars (e.g., Sills et al. 2009 and references therein), which in principle should not be related to the Oosterhoff dichotomy.

We performed Fourier analysis on the RR Lyrae light curves and used the resulting Fourier parameters to calculate physical properties for the stars which will be compared to the properties of RR Lyrae stars in other clusters in a future paper in this series.  We obtained a reddening-corrected distance modulus of $(m-M)_{0}=18.43\pm0.17$ from the RRab and RRc stars.

The average periods for the RRab and RRc stars and the ratio of number of RRc to total number of RR Lyrae all suggest an Oosterhoff-intermediate classification for NGC 1466.  The position of the RRab stars in the Bailey diagram provides for a less clear Oosterhoff determination.

\acknowledgments We wish to thank the referee, J.M. Nemec, for his comments and suggestions which helped improve this paper.  Support for M.C. is provided by Proyecto Fondecyt Regular \#1110326; by Proyecto Basal PFB-06/2007; and by FONDAP Centro de Astrof\'{i}sica 15010003.  M.C. and J.B. are supported by the Ministry for the Economy, Development, and Tourism's Programa Inicativa Cient\'{i}fica Milenio through grant P07-021-F, awarded to The Milky Way Millennium Nucleus.  Support for H.A.S. and C.A.K. is provided by NSF grants AST 0607249 and AST 0707756.

\end{document}